\documentclass[10pt,journal,twocolumn]{IEEEtran}
\usepackage{amsmath,amsfonts,amssymb}
\usepackage{graphicx,cite}
\usepackage{algorithm}
\usepackage{algpseudocode}
\usepackage{bm}
\usepackage{color}
\usepackage{url}
\usepackage{tikz}
\usetikzlibrary{shapes,arrows,positioning}
\usepackage{amsthm}
\usepackage[table]{xcolor}

\newtheorem{theorem}{Theorem}

\theoremstyle{definition}

\usepackage{epstopdf}
\theoremstyle{remark}
\newtheorem{remark}{Remark}
\usepackage{physics}

\usepackage{bbm}
\usepackage{booktabs}
\usepackage{multirow}
\usepackage{pifont}%
\usepackage{caption}
\usepackage{hyperref}
\hypersetup{
    colorlinks=true,
    linkcolor=black,
    filecolor=magenta,      
    urlcolor=cyan,
        }
\usepackage{scalefnt}

\newcommand{\xmark}{\ding{55}}%
\definecolor{lightorange}{RGB}{255,200,120}

\title{Exact Outage Probability and Ergodic Capacity Analysis of NOMA in Rayleigh Fading Channels}
 
\author{Arafat~Al-Dweik,~\IEEEmembership{Senior~Member,~IEEE}, Alok~Kumar~Shukla,~\IEEEmembership{Member,~IEEE}, and {Sami~Muhaidat}, \IEEEmembership{Senior~Member,~IEEE}%
\thanks{Arafat Al-Dweik (e-mail: arafat.dweik@ku.ac.ae, dweik@fulbrightmail.org), Alok Kumar Shukla (e-mail: alok94@ieee.org), and Sami Muhaidat (e-mail: muhaidat@ieee.org), are with the Department of Computer and Information Engineering, Khalifa University,
Abu Dhabi 127788, UAE.}
}

\usepackage[acronym]{glossaries}

\makeglossaries

\newacronym{noma}{NOMA}{non-orthogonal multiple access}
\newacronym{sic}{SIC}{successive interference cancellation}
\newacronym{awgn}{AWGN}{additive white Gaussian noise}
\newacronym{ber}{BER}{bit-error rate}
\newacronym{sinr}{SINR}{signal-to-interference-plus-noise ratio}
\newacronym{isac}{ISAC}{integrated sensing and communication}
\newacronym{bpsk}{BPSK}{binary phase-shift keying}
\newacronym{snr}{SNR}{signal-to-noise ratio}
\newacronym{pdf}{PDF}{probability density function}
\newacronym{cdf}{CDF}{cumulative distribution function}
\newacronym{op}{OP}{outage probability}
\newacronym{ec}{EC}{ergodic capacity}
\newacronym{nu}{NU}{near user}
\newacronym{fu}{FU}{far user}
\newacronym{mc}{MC}{Monte Carlo}
\newacronym{mimo}{MIMO}{multiple-input multiple-output}
\newacronym{csi}{CSI}{channel state information}
\newacronym{qpsk}{QPSK}{quadrature phase shift keying}

\begin{document}
\scalefont{0.96}
\maketitle
\begin{abstract}
This work derives the exact \gls{op} and \gls{ec} for the \gls{nu} in the widely adopted two-user downlink \gls{noma} over fading channels. By noting that the noise and fading become dependent after \gls{sic}, the exact analysis is derived by considering the joint \glspl{pdf} of the post-\gls{sic} noise and fading, which are typically considered to be independent and modeled using the same \glspl{pdf} before the \gls{sic}. The derived exact \glspl{pdf} are used to evaluate the impact of residual interference accurately. The derived interference and noise \glspl{pdf} are used to derive an exact closed-form formula for \gls{nu} outage and a single-integral expression for \gls{ec}. Moreover, a closed-form accurate expression is derived for the \gls{ec}. Unlike existing work, the derived formulae are parameter-free, leading to more accurate performance evaluation of such systems. Monte Carlo simulation results validate the derived analysis and demonstrate that legacy Gaussian/residual-factor models can significantly misestimate outage and \gls{ec} at low–to–moderate \glspl{snr} and under unbalanced power allocation. Moreover, the obtained results show that the widely considered residual interference factor, which is bounded by $[0,1]$, is not sufficient to capture the actual impact of residual interference due to a \gls{sic} failure, and it cannot be treated as an independent variable because it depends on the power allocation, \gls{snr} and outage threshold. In addition to the fading-noise dependence, for two-dimensional modulations, the real and imaginary components of the noise become dependent as well. 
\end{abstract}
\glsresetall
\glsresetall

\section{Introduction}
\Gls{noma} has been extensively investigated for next-generation wireless networks due to its ability to serve multiple users over the same time–frequency resources, thereby improving spectral efficiency \cite{10075436}. In downlink \gls{noma}, the \gls{nu} typically employs \gls{sic} to decode and cancel the \gls{fu} signal before detecting its own. A critical challenge is error propagation because \gls{sic} errors cause residual interference, which can significantly degrade subsequent detection performance. Consequently, performance assessment must go beyond idealized assumptions and account for practical \gls{sic} behavior \cite{Yahya2023Survey}. Beyond spectral‑efficiency gains, reliability under practical deployments and constraints, e.g., randomly deployed receivers and cognitive interference limits, has been investigated in \cite{Zhang2023ReliableNOMA,Bariah2020NOMACognitive}, while recent works highlight that the \gls{sic} stage itself can be a primary limiter or enabler of performance \cite{Ouyang2023NOMAISAC,Semira2021UplinkIOT}.

Most analytical studies of the two-user power-domain \gls{noma} adopt one of two models: perfect \gls{sic}, in which the interferer is always eliminated, or imperfect \gls{sic}, which considers cases in which the \gls{sic} is canceled correctly or incorrectly. For the unsuccessful cancellation, a fixed fraction of the \gls{fu} power is lumped into a Gaussian disturbance via a tunable imperfection factor  \cite{Yahya2023Survey,Yahya2021ExactBER,Aldababsa2020BERNOMA}. Although such models are convenient, they do not capture the statistics induced by hard-decision \gls{sic}. In \gls{awgn} channels, Al-Dweik \textit{et al.} \cite{AlDweik2024WCL} showed that the post-\gls{sic} disturbance is non-Gaussian; rather, it is a truncated-Gaussian mixture whose form depends on whether \gls{sic} succeeds or fails. At the error-rate level, explicitly modeling these distinct post-\gls{sic} regimes tightens analysis and reveals systematic deviations from legacy Gaussian assumptions \cite{AlDweik2024WCL}.

\subsection{Related Work}
\label{sec:lit_review_noma}
The performance characterization of downlink \gls{noma} systems has attracted substantial research attention, with \gls{ber}, \gls{op}, and \gls{ec} serving as primary metrics for evaluating reliability and throughput, respectively. For users who perform \gls{sic} to extract their own data, the performance analysis is conducted under the assumption of perfect or imperfect \gls{sic}. The prefect \gls{sic} assumption aims at deriving tractable closed-form expressions, while imperfect \gls{sic} analysis aims at capturing the impact of error propagation due to the errors performed while canceling the interference of high-power users, at the expense of more tedious derivations and highly intricate expressions. Some examples are summarized below. 
Cui \textit{et al.} \cite{8502922} considered the \gls{op} of a \gls{mimo}-\gls{noma} system where the power allocation and beamforming vectors
are jointly designed to maximize the system utility. To enable mathematical tractability, the perfect \gls{sic} assumption is adopted. 
The work in \cite{8466830} considers the outage achievable rate region of \gls{noma} when all users are required to satisfy their \gls{op} constraints. The \gls{sic} order is performed under the assumption of statistical channel information. The results are obtained for the two-user scenario with Rayleigh fading and perfect \gls{sic} assumption.
Li \textit{et al.} \cite{9031560} derived tractable closed-form expressions for the \gls{op} of the single cell multi-carrier \gls{noma} where statistical \gls{csi} is available only at the transmitter side. The analysis is presented for the two-user Rayleigh-fading scenario, with the perfect \gls{sic} assumption. 

The authors in \cite{8660720} propose a cooperative \gls{noma} system with imperfect \gls{sic} in an underlay cognitive radio network. Considering
that the channel follows a Rayleigh distribution, closed-form formulae of the exact \gls{op} for the secondary users are derived. However, the imperfect \gls{sic} factor is treated as an independent variable regardless of the \gls{snr}.
The \gls{op} of uplink and downlink \gls{noma} is considered \cite{9285248} with a generalized fading channel. The \gls{op} at the users is evaluated for an instantaneous \gls{csi}. The presented analysis considers the perfect \gls{sic} scenario, and thus does not capture the propagation error due to the \gls{sic} errors.
Adaptive user pairing for \gls{noma} With imperfect \gls{sic} is considered in \cite{mouni2021adaptive}. The work derives bounds on power allocation factors and shows the effect of imperfect \gls{sic} on the
rate performance. The work considers the imperfect \gls{sic} factor as an independent variable and evaluates the rate for various values regardless of the \gls{snr}.
In \cite{singh2020outage}, a cooperative \gls{noma}-based relay system is proposed and analyzed. The proposed model allows a source and a
relay to transmit a superimposed signal in the first and second time slots, respectively. The exact closed-form expressions for the \gls{op} are derived in a Rayleigh fading
environment with \gls{sic} at the relay and destination. The residual interference is modeled using the imperfect \gls{sic} coefficient, and the post-\gls{sic} noise and fading are considered independent of the \gls{sic} outcome. 

Table \ref{Tab: papers-summary} summarizes some selected \gls{noma} articles and compares them with this work.
\begin{table*}[tbp] \centering%
\caption{Summary of selected \gls{noma} articles with outage and/or \gls{ec} analysis. The {p.} and {Ip.} stand for perfect and imperfect.}
\label{Tab: papers-summary}%
\begin{tabular}{|c|l|c|c|c|c|c|c|c|}
\hline
\textbf{Ref.} & \textbf{Main Focus} & \textbf{Outage} & \textbf{Capacity} & 
\textbf{p. SIC} & \textbf{Ip. SIC} & \textbf{Ip. SIC Factor} & \textbf{%
Channel} & \textbf{No.} \textbf{Users} \\ \hline\hline
\multicolumn{1}{|c|}{\cite{8502922}} & Power allocation for \gls{mimo} & \multicolumn{1}{|c|}{$\checkmark$} & 
\multicolumn{1}{|c|}{\xmark} & \multicolumn{1}{|c|}{$\checkmark$} & \xmark & 
\multicolumn{1}{|c|}{\xmark} & \multicolumn{1}{|c|}{Rayleigh} & \multicolumn{1}{|c|}{2} \\ 
\hline
\multicolumn{1}{|c|}{\cite{8466830}} & Rate region & \multicolumn{1}{|c|}{\xmark} & 
\multicolumn{1}{|c|}{$\checkmark$} & \multicolumn{1}{|c|}{$\checkmark$} & \xmark & 
\multicolumn{1}{|c|}{\xmark} & \multicolumn{1}{|c|}{Rayleigh} & \multicolumn{1}{|c|}{2} \\ 
\hline
\cite{9031560}& Analysis without \gls{csi}  &$\checkmark$  & \xmark & $\checkmark$ & \xmark & \xmark & Rayleigh & 2 \\ \hline
\cite{8660720}& Underlay cognitive radio & $\checkmark$ & \xmark & \xmark & $\checkmark$ &$ \in[0,1]$  & Rayleigh & 2 \\ \hline
\cite{9285248}& Analysis in generalized channel& $\checkmark$  & \xmark  &$\checkmark$  &\xmark & \xmark & General & General \\ \hline
\multicolumn{1}{|c|}{\cite{mouni2021adaptive}} & Adaptive user pairing & \multicolumn{1}{|c|}{
\xmark} & \multicolumn{1}{|c|}{$\checkmark $} & \multicolumn{1}{|c|}{$%
\checkmark $} & $\checkmark $ & \multicolumn{1}{|c|}{$ \in \left[ 0,1%
\right] $} & \multicolumn{1}{|c|}{Rayleigh} & \multicolumn{1}{|c|}{$2$} \\ 
\hline
\cite{singh2020outage}& Cooperative relaying & $\checkmark$ & \xmark & \xmark & $\checkmark$ & $\in[0,1]$ &Rayleigh  & 2 \\ \hline

\cite{alqahtani2021}& Performance Analysis & $\checkmark$ & \checkmark & $\checkmark$ & \xmark & \xmark & $\alpha$-$\mu$  & General \\ \hline

\cellcolor{lightorange}This work&\cellcolor{lightorange} Exact Analysis &\cellcolor{lightorange} $\cellcolor{lightorange}\checkmark$ &\cellcolor{lightorange} $\checkmark$ &\cellcolor{lightorange} \xmark &\cellcolor{lightorange} $\checkmark$ &\cellcolor{lightorange} \xmark &\cellcolor{lightorange}Rayleigh  &\cellcolor{lightorange} 2 \\ \hline
\end{tabular}%
\end{table*}%

\subsection{Motivation and Contributions}
As can be noted from the surveyed literature, which is partially discussed in this article, the impact of \gls{sic} in \gls{noma} is limited to the presence or absence of residual interference. That is, if the \gls{sic} process succeeded, then there is no residual interference. Otherwise, there is residual interference that affects the detection process for users who need to perform \gls{sic}. Nonetheless, the \gls{sic} process affects the noise and fading characteristics. Based on an extensive literature search, and to the best of the authors' knowledge, there is no work that characterizes the noise and fading after \gls{sic}, apart from \cite{AlDweik2024WCL}, which considers the \gls{awgn} channel model. Consequently, the main contributions of this work are:
\begin{enumerate}
    \item Evaluates the impact of the \gls{sic} on the noise and fading, which is necessary to evaluate the performance of users who need to perform \gls{sic}. In particular, the work shows that noise and fading after \gls{sic} become dependent, and that their \glspl{pdf} differ from those before \gls{sic}. 

    \item The residual interference is modeled accurately, where the impact of the unsuccessful interference cancellation may increase the interference power.

    \item Demonstrate that the \gls{sic} imperfection factor depends on the \gls{snr}, power allocation, and outage threshold, and thus should not be treated as an independent variable, which is the case in the majority of existing work.
    
    \item Prove that the typical range of the imperfection factor $[0,1]$ is not sufficient to capture the impact of imperfect \gls{sic} for outage analysis, and should be set to zero for \gls{ec} analysis. 
    
    \item Derives the joint \gls{pdf} of the noise and fading after \gls{sic}, and uses it to derive the marginal \glspl{pdf}, which are used to derive exact expressions for the \gls{op} and \gls{ec} for strong \gls{noma} users. 
    
    \item Simulation results verify the accuracy of the derived expressions while highlighting the divergence regions of conventional imperfect‑\gls{sic} models. 
\end{enumerate}

\subsection{Terminology and Notation}
Throughout this paper, the terminology used is as follows. A successful \gls{sic} indicates that the \gls{sic} process was correct for a particular symbol, while a \gls{sic} failure corresponds to an incorrect \gls{sic}. A perfect \gls{sic} indicates the assumption where all \gls{sic} processes are successful. In contrast, imperfect \gls{sic} indicates cases in which the analysis considers both successful and failed \gls{sic} processes. The term legacy refers to schemes in which the noise and fading statistics remain unchanged before and after \gls{sic}, and the imperfect \gls{sic} is modeled using a residual interference factor. 

The conditional \gls{pdf} $f_{N|\mathcal{S}}(n)$ is expressed as $f_{W}(w)$ and
$f_{N|\mathcal{F}}(n)$ is expressed as $f_{Z}(z)$. The conditional noise and fading are respectively given by $N|\mathcal{S}\rightarrow W$ and 
$N|\mathcal{F}\rightarrow Z$. The statistical average is denoted by $\mathbb{E}[\cdot]$.

\subsection{Paper Organization}
The remainder of the paper is organized as follows:
Sec. \ref{sec:sys_model} presents the \gls{noma} system model and legacy analysis. 
Sec. \ref{sec:postSIC} derives the \glspl{pdf} of the noise and fading after successful and failure \gls{sic}. 
Sec. \ref{sec:SINR_SNR_OP} derives the post-\gls{snr}/\gls{sinr} and the exact \gls{op}. 
Sec. \ref{sec: EC_Analysis} presents the \gls{ec} analysis with successful and failure \gls{sic}. 
Sec. \ref{sec:QPSK} derives the post-\gls{sic} \gls{pdf} of the fading and noise using \gls{qpsk} with successful \gls{sic}. 
Monte Carlo simulation results and numerical analytical results are discussed in Sec. \ref{sec:num}, and finally, Sec. \ref{sec:conclusion} concludes the article.

\section{NOMA System Model and Legacy Analysis}
\label{sec:sys_model}
This work considers the widely adopted two-user downlink \gls{noma}, in which a base station transmits a superposition of the two users’ symbols, which can be written as
\begin{equation}
\label{eq:X}
X = \sqrt{\alpha_1} s_1 + \sqrt{\alpha_2} s_2
\end{equation}
where $\alpha_1$ and $\alpha_2$ are the powers of user $U_1$ (\gls{fu}) and $U_2$ (\gls{nu}), respectively, where $\alpha_1 + \alpha_2 = 1$. For reliable \gls{sic} detection, it is typically considered that $\alpha_1 > \alpha_2$. Both users are considered to employ \gls{bpsk} modulation, and thus $\{s_1, s_2\} \in \{-1, +1\}$. 
The exact \gls{op} and \gls{ec} for $U_1$ are already derived in the literature, and thus not considered in this work. The received signal at $U_2$ is given by
\begin{equation}
\label{eq:R2}
y_2 = h_2 X + N_2
\end{equation}
where $h_2$ is the complex channel gain and $N_2 \sim \mathcal{CN}(0, 2\sigma_{\mathrm{n}}^2)$ is a circularly symmetric \gls{awgn}. After, channel phase, $\theta_2=\arg{\{h_2\}}$, estimation and compensation we obtain,
\begin{equation}
Y_2 = \mathrm{e}^{-j\theta}y_2=\beta_2 (\sqrt{\alpha_1} s_1 + \sqrt{\alpha_2} s_2) + N_2
\label{eq:Y}
\end{equation}
where $\beta_2 = |h_2|$ and the \gls{awgn} $N_2$ notation remains unchanged because its \gls{pdf} is phase invariant. Since the subsequent analysis focuses exclusively on $U_2$, the user index is omitted hereafter unless explicitly required for clarity.

To detect its own signal, $U_2$ first detects $s_1$ while treating $s_2$ as unknown interference. For \gls{bpsk}, the desired signal component is real, and thus, the detection is a simple sign decision on $Y\leftarrow \Re(Y)$. Then, the estimated symbol $\hat{s}_1$ is used to cancel the interference of $U_1$. The signal after \gls{sic} is given by,
\begin{equation} 
\label{eq:SIC_subtraction}
\acute{Y} = Y - \beta \sqrt{\alpha_1} \hat{s}_1.
\end{equation}
 

\subsection{Legacy Imperfect-SIC Analysis}\label{subsec:legacy}

In the literature, the received signal of $U_2$ after \gls{sic} is given by \cite[Eq. (3)]{10891917},
\begin{equation}
\label{eq:sic_w}
\acute{Y} = \beta\sqrt{\alpha_2} s_2 + \zeta \beta\sqrt{\alpha_1}s_1  +N
\end{equation}
where $\eta \in [0,1]$ is the \gls{sic} imperfection factor,
$\zeta=0$ corresponds to a successful \gls{sic}, $\hat{s}_1 = s_1$, with full interference removal, whereas larger $\eta$ values model stronger residual interference from the \gls{fu} signal when $\hat{s}_1 \neq s_1$. It is worth noting that the models in \eqref{eq:SIC_subtraction} and \eqref{eq:sic_w} are not identical. The interference in \eqref{eq:SIC_subtraction} is $\beta\sqrt{\alpha_1}(s_1-\hat{s}_1)$ while in \eqref{eq:sic_w} it is given by $\eta \beta\sqrt{\alpha_1} s_1$. The difference between the two models is that in \eqref{eq:SIC_subtraction} it is assumed that the \gls{sic} is applied regardless of the detection result of $s_1$ while in \eqref{eq:sic_w}, it is assumed that \gls{sic} is applied only if $\hat{s}_1=s_1$. Although such a justification is rarely mentioned in existing work \cite{10891917}, it is the most reasonable justification for such an assumption. However, it could be practically challenging to determine whether the \gls{sic} is successful or not for each symbol. 

Due to the absence of knowledge about the \gls{sic} outcome, the receiver proceeds to decode $s_2$ regardless of the \gls{sic} outcome. In the literature, the noise term $N$ in \eqref{eq:R2} and \eqref{eq:sic_w} are considered identically distributed regardless of the value of $\eta$. 
Therefore, based on \eqref{eq:sic_w}, the legacy \gls{nu} instantaneous \gls{sinr} is given by \cite{8309422, 8660720, 9285248, singh2020outage},
\begin{equation}
\label{eq:snr-1}
\gamma
= \frac{\alpha_2 \beta^2}{\zeta\,\alpha_1 \beta^2+\sigma_{\mathrm{n}}^2}.
\end{equation}
where $\zeta=\eta^2$. In \eqref{eq:snr-1}, the noise power is $\sigma^2_{\mathrm{n}}$ because only the real part of the noise is considered with \gls{bpsk}. For two-dimensional modulation schemes such as \gls{qpsk}, it should be $2\sigma^2_{\mathrm{n}}$.  
\begin{remark}
    The imperfect \gls{sic} factor $\eta$ does not accurately capture the residual interference after a failure \gls{sic}. For example, given that \gls{bpsk} is used in \eqref{eq:SIC_subtraction}, then $\acute{Y}$ after a \gls{sic} failure can be written as $\acute{Y}=\beta \sqrt{\alpha_2}s_2-2\beta \sqrt{\alpha_1}s_1$. Consequently, when $\alpha_1$ is small, the \gls{sic} error probability increases, and thus, using the maximum possible value of $\zeta=1$ would not be sufficient to capture the actual interference. 
\end{remark}

Based on \eqref{eq:snr-1}, \gls{op} in Rayleigh fading is given by,
\begin{equation}
\label{eq:PO_legacy}
\!\!\!P_{{O}}=
\begin{cases}
1-\exp\!\Big(\dfrac{-\gamma_{\rm th}/\bar{\gamma}}{\alpha_2-\zeta\alpha_1\gamma_{\rm th}\,}\Big), & \alpha_2>\zeta\alpha_1\gamma_{\rm th},\\[0.4em]
1, & \text{otherwise.}
\end{cases}
\end{equation}
where $\gamma_{\rm th} \triangleq 2^{R} - 1$ and $R$ denotes the target rate of $U_2$, $\bar{\gamma}=\Omega/\sigma^2_{\mathrm{n}}$ and $\Omega=\mathbb{E}(\beta^2)=2\sigma^2_{\beta}$, and
\begin{equation}
\label{eq:Rayleigh}
f_\beta(\beta) = \frac{2\beta}{\Omega} \mathrm{e}^{\frac{-\beta^2}{\Omega}}, \hspace{5mm} \beta \ge 0.
\end{equation}

The \gls{ec} is given by
\begin{align}
\label{eq:C_lgacy}
C& =\mathbb{E}_{\beta }\!\big[\log _{2}\!\big(1+\gamma \big)\big]  \notag \\
& =\int_{0}^{\infty }\log _{2}\!\Big(1+\frac{\alpha _{2}\beta ^{2}}{\zeta
\,\alpha _{1}\beta ^{2}+\sigma _{\mathrm{n}}^{2}}\Big)f_{\beta }(\beta
)\,d\beta   \notag \\
& =\frac{1}{\ln 2}\!\mathrm{e}^{\frac{1/\bar{\gamma}}{\zeta \alpha
_{1}+\alpha _{2}}}E_{1}\!\Big(\frac{1/\bar{\gamma}}{\zeta \alpha _{1}+\alpha
_{2}}\Big)-\mathrm{e}^{\frac{1/\bar{\gamma}}{\zeta \alpha_{1}}}E_{1}\!\Big(%
\frac{1/\bar{\gamma}}{\zeta \alpha _{1}}\Big)
\end{align}
where $E_1(x)\!\! \triangleq\!\! \int_x^{\infty} \frac{\mathrm{e}^{-t}}{t}\,dt$ is the exponential integral function. 
\begin{remark}
   In \eqref{eq:PO_legacy} and \eqref{eq:C_lgacy}, the residual interference parameter $\zeta$ is considered as an independent system variable, and thus, the system performance is evaluated for various values of $\zeta$ in the range $[0,1]$ \cite{8309422, 8660720, 9285248, singh2020outage}. However, as will be shown in this work, the noise variance $\sigma^2_{\mathrm{n}}$ and the \gls{pdf} of $\beta$ depend on $\zeta$. Consequently, the instantaneous \gls{snr} in \eqref{eq:snr-1} becomes invalid because the noise power is not $\sigma^2_{\mathrm{n}}$ and the channel gain $\beta$ is not Rayleigh. More importantly, the noise and fading may become dependent and can not be considered independent, as in the case when \gls{op} is evaluated for $U_1$.
\end{remark}

\section{Exact Post-SIC Noise and Fading Characterization using BPSK}
\label{sec:postSIC}
As demonstrated in \cite{AlDweik2024WCL}, the post-\gls{sic} noise at $U_2$ is conditioned on the \gls{sic} outcome. Specifically, the initial \gls{awgn} $N \sim \mathcal{N}(0, \sigma_{\mathrm{n}}^2)$ is transformed into a non-Gaussian random variable $W$ upon successful \gls{sic} ($\mathcal{S}$), or $Z$ in the event of failure ($\mathcal{F}$). Similarly, the distribution of $\beta$ is also outcome-dependent, denoted as $\beta_{\mathcal{S}}$ and $\beta_{\mathcal{F}}$ for the success and failure cases, respectively.
It is worth noting that the distribution of $\beta$ is also conditioned on the outcome of \gls{sic}. Accordingly, it is denoted as $\beta_{\mathcal{S}}$ and $\beta_{\mathcal{F}}$ for the cases of successful and failed \gls{sic}, respectively.


\subsection{Detection Thresholds and SIC Success Probability}

Using \gls{bpsk}, the constellation points of $X$ in \eqref{eq:X} are defined as $\{X_{00},X_{10}, X_{01}, X_{11}\}$, in $X_{ij}$, $i$ and $j$ are the bit values of $U_1$ and $U_2$, respectively. Therefore, $X_{ij}=\bar{i}\sqrt{\alpha_1}+\bar{j}\sqrt{\alpha_2}$ where $\bar{i}=2i-1$ and $\bar{j}=2j-1$. By defining $A_{ij}=\beta X_{ij}$, then for a fixed pair $(s_1,s_2)$, \gls{sic} succeeds if $\hat{s}_1 = s_1$, which defines the following noise intervals for $N$,
\begin{align*}
X_{11}: 
Y \ge 0 \;\Leftrightarrow\; N \ge  -\beta X_{11}, \nonumber\\
X_{10}:
Y \ge 0 \;\Leftrightarrow\; N \ge -\beta X_{10}, \nonumber\\
X_{01}:Y < 0 \;\Leftrightarrow\; N < \beta X_{10}, \nonumber\\
X_{00}:
Y < 0 \;\Leftrightarrow\; N <  \beta X_{11}.
\label{eq:thresholds}
\end{align*}
The conditional \gls{sic}-success probabilities for the four points follow from Gaussian tail probabilities, which is given by 
\begin{equation}
 \Pr(\mathcal{S} \mid X_{ij}, \beta) = p_{\mathcal{S}}|\beta=\Phi\left( \frac{\beta\sqrt{\alpha_1} + \bar{i}~\bar{j} \beta\sqrt{\alpha_2}}{\sigma_{\mathrm{n}}} \right)
\end{equation}
For Rayleigh fading, see Appendix \ref{appndx-A} , the probability $s_1$ correct detection conditioned on the transmitted \gls{noma} symbol is given by,  
\begin{equation}
\label{eq:ps-1}
p_{\mathcal{S}}=\frac{1}{2}+\frac{1}{2}\sqrt{\frac{X_{ij}^{2}}{X_{ij}^{2}+\frac{2}{%
\bar{\gamma}}}}, \hspace{5mm}\bar{\gamma}=\frac{\Omega}{\sigma^2_{\mathrm{n}}}.
\end{equation}
Therefore, $p_{\mathcal{F}}=1-p_{\mathcal{S}}$.

\subsection{Post-SIC Noise and Fading PDFs}
\begin{figure*}[t]
    \centering
    \includegraphics[scale=0.9]{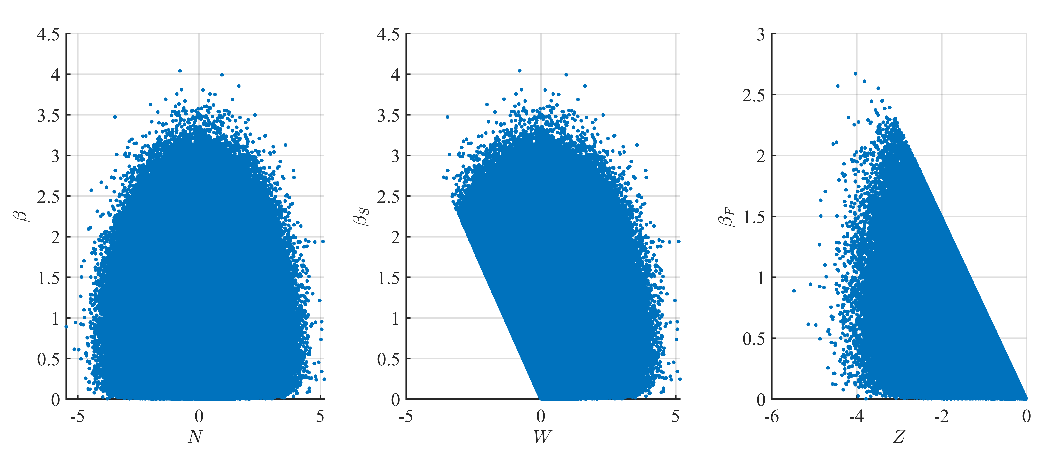}
    \caption{Scatter plots of the conditional and unconditional noise and fading with \gls{snr} of $0$ dB,  $\alpha_1=0.8$, and $X=X_{11}$.}
    \label{fig:scatt_0}
\end{figure*}
\begin{figure*}[t]
    \centering
    \includegraphics[scale=0.9]{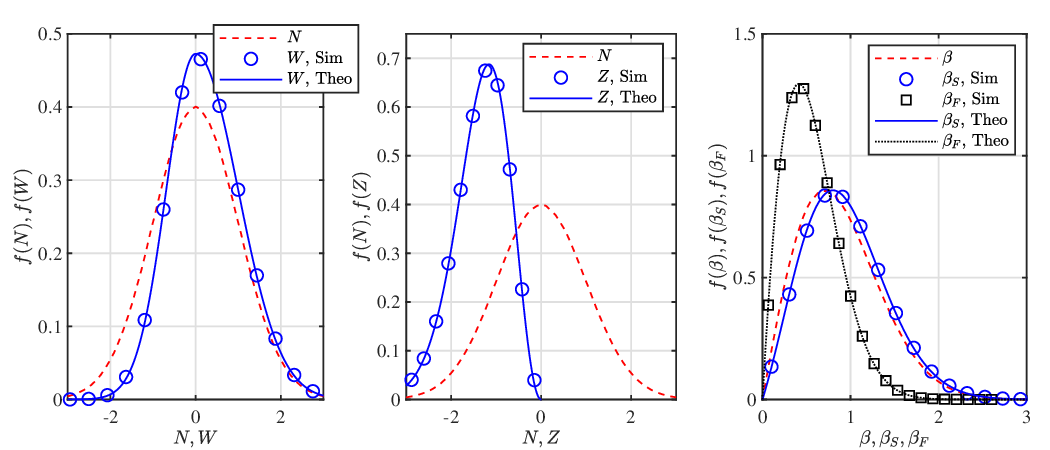}
    \caption{The conditional and unconditional \glspl{pdf} of $N$ and $\beta$ with \gls{snr} of $0$ dB,  $\alpha_1=0.8$, and $X=X_{11}$.}
    \label{fig:cod_PDFs_0}
\end{figure*}
\begin{figure*}[t]
    \centering
    \includegraphics[scale=0.9]{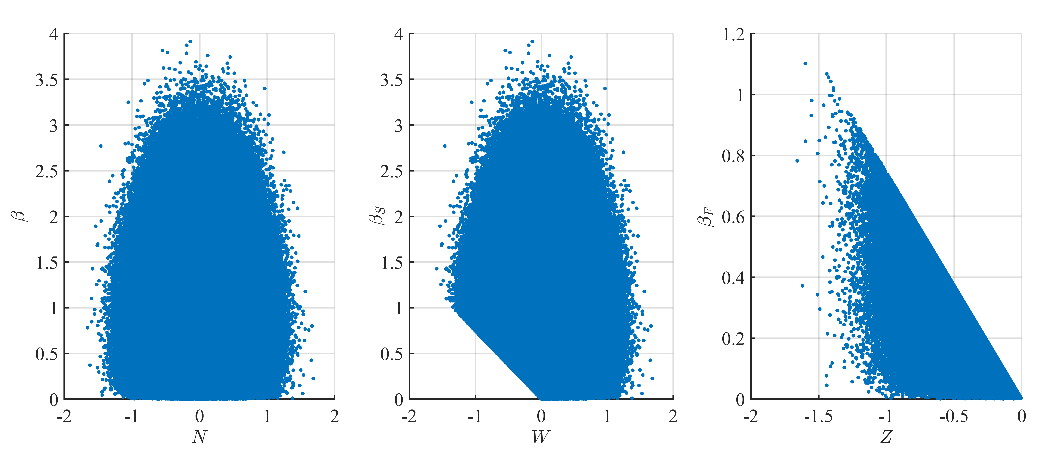}
    \caption{Scatter plots of the conditional and unconditional noise and fading with \gls{snr} of $10$ dB,  $\alpha_1=0.8$, and $X=X_{11}$.}
    \label{fig:scatt_10}
\end{figure*}
\begin{figure*}
    \centering
    \includegraphics[scale=0.9]{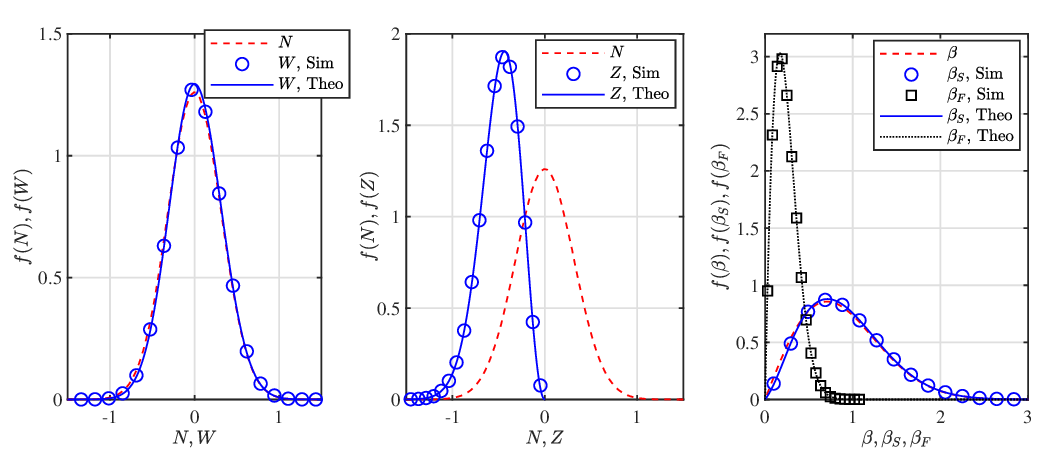}
    \caption{The conditional and unconditional \glspl{pdf} of $N$ and $\beta$ with \gls{snr} of $10$ dB,  $\alpha_1=0.8$, and $X=X_{11}$.}
    \label{fig:cod_PDFs_10}
\end{figure*}

Based on \eqref{eq:SIC_subtraction}, the post-\gls{sic} signal for $U_2$ can be written as
\begin{equation}
    \acute{Y} = \beta\sqrt{\alpha_2}\,s_2 + \beta\sqrt{\alpha_1}\bigl(s_1-\hat{s}_1\bigr) + N
\end{equation}
where the second term $ \beta\sqrt{\alpha_1}\bigl(s_1-\hat{s}_1\bigr) \triangleq D$, 
is the residual \gls{fu} interference and the \gls{awgn} sample $N\sim\mathcal{N}(0,\sigma_{\mathrm{n}}^2)$.  For \gls{bpsk} with hard-decision detection, an erroneous decision on $s_1$ flips its sign, i.e., $\hat{s}_1=-s_1$, and thus $s_1-\hat{s}_1 = 2s_1$. Consequently, 
the post-\gls{sic} signal conditioned on the \gls{sic} outcome can be written as
\begin{equation}
\label{eq:YS}
   \acute{Y}_{\mathcal{S}} = \beta_\mathcal{S}\sqrt{\alpha_2}\,s_2 + W
\end{equation}
\begin{equation}
\label{eq:YF}
 \acute{Y}_\mathcal{F} =   \beta_{\mathcal{F}}\sqrt{\alpha_2}\,s_2 + 2\beta_{\mathcal{F}}\sqrt{\alpha_1}s_1 + Z.
\end{equation}
As can be noted from \eqref{eq:YS} and \eqref{eq:YF}, the conditioning on the \gls{sic} outcome introduced the following transformations: $N|\mathcal{S}\rightarrow W$, $N|\mathcal{F}\rightarrow Z$, $\beta|\mathcal{S}\rightarrow \beta_{\mathcal{S}}$, and $\beta|\mathcal{F}\rightarrow \beta_{\mathcal{F}}$. For a fixed $\beta$, the \glspl{pdf} of $W$ and $Z$ are given in \cite{AlDweik2024WCL}, while the \gls{pdf} for a random $\beta$ has never been derived. Moreover, although $\beta$ and $N$ are independent, the condition on the \gls{sic} outcome transforms them to become jointly dependent. The joint and marginal \glspl{pdf} are given in Theorem \ref{Theo:PDFs-1} for the case of Rayleigh fading, that is

\begin{theorem}\label{Theo:PDFs-1}
    Given that the transmitted \gls{noma} symbol $X\in\{X_{11}, X_{10}\}$, then $X_{ij}=\sqrt{\alpha_1}\pm \sqrt{\alpha_2}>0$. After a successful \gls{sic}, the noise and fading, $W$ and $\beta_{\mathcal{S}}$, become dependent with the following joint and marginal \glspl{pdf},
\begin{equation}
   \!\!\! f_{\beta_{\mathcal{S}},W}(\beta_{\mathcal{S}},w) = \frac{f_N(w)  f_{\beta}(\beta_{\mathcal{S}})}{p_{\mathcal{S}}}, \beta_{\mathcal{S}}\geq 0, w\geq -X_{ij}\beta_{\mathcal{S}}
\end{equation}
the marginal \gls{pdf} of $\beta_{\mathcal{S}}$, 
\begin{equation}
\label{eq:f_BS_1}
f_{\beta_{\mathcal{S}}}(\beta _{\mathcal{S}})=\frac{2\beta _{\mathcal{S}}}{%
\Omega p_{\mathcal{S}}}\Phi \left( \frac{|X_{ij}|\beta_{\mathcal{S}}}{\sigma _{%
\mathrm{n}}}\right) \mathrm{e}^{-\frac{\beta_{\mathcal{S}}^{2}}{\Omega }%
},\quad \beta_{\mathcal{S}} \geq 0
\end{equation}
and the marginal \gls{pdf} of $W$ 
 \begin{equation}
 \label{eq:fW_1}
f_{W}(w)=%
\begin{cases}
\frac{1}{p_{\mathcal{S}}}f_{N}\left( w\right)  & w\geq 0 \\[5pt]
\frac{f_{N}\left( w\right) }{p_{\mathcal{S}}}\exp \left( \frac{-w^{2}}{%
X_{ij}^{2}\Omega }\right)  & w<0%
\end{cases}%
\end{equation}

\title{Derivation of Conditional PDFs for the Error Case ($Y < 0$)}
\author{Statistical Analysis of SIC Environments}
\date{\today}
\maketitle
The joint \gls{pdf} when the \gls{sic} fails
\begin{equation}
\!\!\!f_{\beta_{\mathcal{F}},Z}(\beta _{\mathcal{F}},z)= \frac{f_N(z)  f_{\beta}(\beta_{\mathcal{F}})}{p_{\mathcal{F}}},\\
\beta_{\mathcal{F}} \geq 0, z < -X_{ij}\beta_{\mathcal{F}} 
\end{equation}
where $p_{\mathcal{F}}=1-p_{\mathcal{S}}$.

The Marginal \gls{pdf} of $\beta _{\mathcal{F}}$, 
\begin{equation}
\label{eq:f_BF_1}
f_{\beta_{\mathcal{F}}}(\beta _{\mathcal{F}})=\frac{2\beta _{\mathcal{F}}}{%
\Omega~ p_{\mathcal{F}}}\mathrm{e}^{-\frac{\beta _{\mathcal{F}}^{2}}{\Omega }%
}Q\left( \frac{|X_{ij}|\beta _{\mathcal{F}}}{\sigma_{\mathrm{n}}}\right) ,\quad \beta _{\mathcal{F}}\geq 0
\end{equation}
and the marginal \gls{pdf} of $Z$,
\begin{equation}
\label{eq:fZ_1}
    f_{Z}(z) = \frac{f_N(z)}{p_{\mathcal{F}}} \left[ 1 - \exp\left( -\frac{z^2}{X_{ij}^2\Omega} \right) \right], \quad z < 0.
\end{equation}
 \begin{IEEEproof}
        The proof is given in Appendix \ref{appndx-A}.
    \end{IEEEproof}
\end{theorem}
For the other two cases of $X\in\{X_{00}, X_{01}\}$, they can be obtained via the constellation symmetry. For $p_{\mathcal{S}}$, it can be computed using \eqref{eq:ps-1} $\forall \{i,j\}$, and similarly  $f_{\beta_{\mathcal{S}}}(\beta_{\mathcal{S}})$  and $f_{\beta_{\mathcal{F}}}(\beta_{\mathcal{F}})$ can be computed using \eqref{eq:f_BS_1} and \eqref{eq:f_BF_1}, respectively. The \glspl{pdf} $f_W(w)$ and $f_Z(z)$ can be computed using \eqref{eq:fW_1} and  \eqref{eq:fZ_1}, respectively, with flipped intervals for $w$ and $z$, e.g., $\{w,z\}>0 \rightarrow \{w,z\}<0$.
\begin{remark}
    The mutual interaction of the noise and fading can be clearly noted from the marginal \glspl{pdf} where  $f_{\beta_{\mathcal{S}}}(\beta_{\mathcal{S}})$ and $f_{\beta_{\mathcal{F}}}(\beta_{\mathcal{F}})$ include the \gls{awgn} variance $\sigma^2_{\mathrm{n}}$. Similarly, $f_W(w)$ and $f_Z(z)$ include the average fading gain $\Omega$. Moreover, both the fading and noise depend on the transmitted symbol $X_{ij}$.
\end{remark}

To illustrate the dependence between the noise and fading, Fig. \ref{fig:scatt_0} presents the scatter plot of a) $\{\beta, N\}$, b) $\{\beta_{\mathcal{S}}, W\}$, and c) $\{\beta_{\mathcal{F}}, Z\}$ for $X=X_{11}$ and $\bar{\gamma}=0$ dB. For $\{\beta, N\}$, the two variables are independent, and thus there are no constraints on their values. For $\beta_{\mathcal{S}}$, the scatter pattern is truncated such that $w>-X_{11}\beta_{\mathcal{S}}$. The truncation is also experienced for $\{\beta_{\mathcal{F}}, Z\}$ where $z<-X_{11}\beta_{\mathcal{F}}$. Obviously, scatter plots b) and c) complement each other. Similarly, Fig. \ref{fig:scatt_10} presents the scatter plots with $\bar{\gamma}=10$ dB, where the same trends can be noted. However, note that the intervals in the scatter plot differ. For \gls{snr} of $10$ dB, the noise is generally small, so for the \gls{sic} to be successful, the values of $\beta_{\mathcal{S}}$ must be large. On the contrary, for the \gls{sic} to fail, the values of $\beta_{\mathcal{F}}$ should be small. 

Fig. \ref{fig:cod_PDFs_0} presents the analytical and simulated \glspl{pdf} for the same cases considered for the scatter plots. As can be seen from $f_{\beta_{\mathcal{F}}}(\beta_{\mathcal{F}})$, the \gls{sic} errors are mostly due to the large values of the noise samples, therefore, high  $\beta_{\mathcal{F}}$ values can still be observed. Nevertheless, small ${\beta_{\mathcal{F}}}$ values can still cause \gls{sic} errors. Therefore, $f_{\beta_{\mathcal{F}}}(\beta_{\mathcal{F}})$ has more weight towards the y-axis. 
When the \gls{sic} is successful, the noise samples are generally small, and thus, the large values of $\beta_{\mathcal{S}}$ have a higher probability. However, the difference is not significant compared to the unconditional Rayleigh. For the case of $\bar{\gamma}=10$ dB, a \gls{sic} failure only happens at deep fading, making $f_{\beta_{\mathcal{F}}}(\beta_{\mathcal{F}})$ more dense near the y-axis. For $\beta_{\mathcal{S}}$, the small noise samples enabled passing all values of $\beta$, making $f_{\beta_{\mathcal{S}}}(\beta_{\mathcal{S}})\approx f_{\beta}(\beta)$. The impact of the noise on the $\beta_{\mathcal{S}}$ and $\beta_{\mathcal{F}}$ can be noted from \eqref{eq:f_BS_1} and \eqref{eq:f_BF_1} where $\sigma_{\mathrm{n}}$ is a factor in both formulae.

The post-\gls{sic} noise \glspl{pdf} are significantly affected by the conditioning. In particular, the noise is not Gaussian anymore, and $\{\mathbb{E}(W), \mathbb{E}(Z)\} \neq 0$. The deviation from the Gaussian distribution is significant for the case of failure \gls{sic} and low \glspl{snr}. For the case of $\bar{\gamma}=10$ dB, $f_W(w)\approx f_N(w)$, while $f_Z(z)$ is significantly different from $f_N(w)$. The impact of the fading on the $W$ and $Z$ can be noted from \eqref{eq:fW_1} and \eqref{eq:fZ_1} where $\Omega$ is a factor in both formulae. For stationary channels, $\Omega$ is fixed, making $f_Z(z)$ and $f_W(w)$ independent of the average channel gain. 

\section{Post-SIC SNR, SINR, and Outage Probability}
\label{sec:SINR_SNR_OP}
The instantaneous \gls{snr} and \gls{sinr} for the $\{\mathcal{S},\mathcal{F}\}$ scenarios are derived as follows:

\subsection{Successful SIC}
In the case of successful \gls{sic}, the interference is canceled, and thus the \gls{snr} is expressed as,
\begin{equation}
\label{gamma_s_1}
    \gamma_{\mathcal{S}}
    = \frac{\alpha_2 \beta_{\mathcal{S}}^2}{\mathbb{E}[W^2]}.
\end{equation}
As shown in \eqref{gamma_s_1}, the second moment of the noise is used instead of the variance because the noise mean is not necessarily zero. The second moment of $w$ can be evaluated as
\begin{equation}
\label{eq:sec_mom_w_1}
\mathbb{E}\left[W^{2}\right] =\int_{-\infty }^{\infty }w^{2}f_{W}(w)dw.
\end{equation}
By substituting \eqref{eq:fW_1} into \eqref{eq:sec_mom_w_1} and evaluating the integral,
\begin{equation}
\label{eq:sec_mom_w_2}
\mathbb{E}\left[W^{2}\right]=\frac{\sigma _{\mathrm{n}}^{2}}{2p_{\mathcal{S}}}\left( 1+|{X}_{ij}|^{3}%
\left[ \frac{\bar{\gamma}}{\bar{\gamma}\,{X}_{ij}^{2}+2}\right] ^{\frac{3}{2}%
}\right).
\end{equation}
Based on \eqref{eq:sec_mom_w_2}, it can be noted that $\mathbb{E}\left[W^{2}\right]< \sigma _{\mathrm{n}}^{2}$ and $\lim_{\bar{\gamma}\rightarrow \infty}=\sigma _{\mathrm{n}}^{2}$. Such behavior is expected because the set of samples that corresponds to a successful \gls{sic} is generally less noisy compared to the original unsampled noise. 

\begin{remark}
    In the literature, e.g.  \cite{alqahtani2021, 9285248}, the perfect \gls{sic} assumption is based on approximating exact performance by setting $\zeta=0$ in \eqref{eq:snr-1}, which implies that the residual error is discarded. However, 
    As can be seen from \eqref{gamma_s_1} and \eqref{eq:sec_mom_w_2}, the typical perfect \gls{sic} approach imposes two other overlooked approximations, which are the channel fading gain and noise after \gls{sic} retain the same \glspl{pdf} before \gls{sic}. 
\end{remark}

\Gls{op} after a successful \gls{sic} can be computed based on \eqref{gamma_s_1}. By defining $\frac{\alpha_2}{\mathbb{E}\left[W^{2}\right]}\triangleq\psi$, then
\begin{align}
\label{PO_S_BPSK_1}
P_{O_{\mathcal{S}}} &=\Pr \left( \gamma _{\mathcal{S}}<\gamma _{\mathrm{%
th}}\right)  =\Pr \left( \psi \beta _{\mathcal{S}}^{2}<\gamma _{\mathrm{th}}\right)  \nonumber \\
&=\Pr \left( \beta _{\mathcal{S}}<\sqrt{\frac{\gamma _{\mathrm{th}}}{\psi }}%
\right) \nonumber \\
&=\int_{0}^{\sqrt{\frac{\gamma _{\mathrm{th}}}{\psi }}}f_{\beta _{\mathcal{S%
}}}\left( \beta _{\mathcal{S}}\right) d\beta _{\mathcal{S}}.
\end{align}
 Substituting $f_{\beta_{\mathcal{S
}}}\left( \beta _{\mathcal{S}}\right )$ in \eqref{eq:f_BS_1} and evaluating the integral gives, 
\begin{multline}
P_{O_{\mathcal{S}}}=\frac{1}{p_{\mathcal{S}}}\left[ \frac{1}{2}-\mathrm{e}^{\frac{-\epsilon ^{2}}{%
\Omega }}\left[ 1+Q\left( \epsilon \sqrt{\dfrac{\bar{\gamma}{X}_{ij}^{2}}{%
\Omega }}\right) \right] \right.\\
\left.+\varphi \Phi \left( \epsilon \sqrt{\dfrac{\bar{%
\gamma}{X}_{ij}^{2}+2}{\Omega }}\right) \right] 
\end{multline}
where $\varphi =\sqrt{\frac{{X}_{ij}^{2}\bar{\gamma}}{{X}_{ij}^{2}\bar{\gamma}%
+2}}$ and $\epsilon=\sqrt{\frac{\gamma _{\mathrm{th}}}{\psi }}$.

\subsection{Unsuccessful SIC}
In the case of \gls{sic} failure, the interference will be present, and thus
\begin{equation} 
    \gamma_{\mathcal{F}}
    = \frac{\alpha_2 \beta_{\mathcal{F}}^2}{4\alpha_1 \beta_{\mathcal{F}}^2 + \mathbb{E}\left[Z^{2}\right]}
        \label{eq:gammaf_cap}
\end{equation}
where the constant $4$ represents the interference power due to the \gls{sic}  failure, i.e., $|2s_1|^2=4$. Similar to the case of $\gamma_{\mathcal{S}}$, the second moment is used to capture the conditional noise mean bias. By noting that $f_Z(z)$ in \eqref{eq:fZ_1} is one-sided, the second moment is computed as, 
\begin{align}
\mathbb{E}\left[Z^{2}\right]  &=\int_{-\infty }^{0}z^{2}f_{Z}(z)dz 
\nonumber \\
&=\frac{\sigma_{\mathrm{n}}^{2}}{2p_{\mathcal{F}}}\left[ 1-\left\vert {X_{ij}}%
\right\vert ^{3}\left[ \frac{\bar{\gamma}}{2+\bar{\gamma}\left\vert {X_{ij}}%
\right\vert ^{2}}\right] ^{\frac{3}{2}}\right].
\end{align}
Therefore, 
\begin{align}
P_{O_{\mathcal{F}}} &=\Pr \left( \gamma _{\mathcal{F}}<\gamma _{\mathrm{th}%
}\right)   \nonumber \\
&=\Pr \left( \frac{\alpha _{2}\beta _{\mathcal{F}}^{2}}{4\alpha _{1}\beta _{%
\mathcal{F}}^{2}+\mathbb{E}\left( Z^{2}\right) }<\gamma _{\mathrm{th}%
}\right)   \nonumber \\
&=\Pr \left( \beta _{\mathcal{F}}^{2}<\frac{\gamma _{\mathrm{th}}\mathbb{E}%
\left( Z^{2}\right) }{\alpha _{2}-4\alpha _{1}\gamma _{\mathrm{th}}}\right). 
\label{eq:POF-1}
\end{align}%
Because $\beta_{\mathcal{F}}^{2}\geq 0$, then from (\ref{eq:POF-1}),%
\begin{equation}
\label{eq:POF-3}
P_{O_{\mathcal{F}}}=\left\{ 
\begin{array}{lc}
1, & \alpha _{2}<4\alpha _{1}\gamma _{\mathrm{th}} \\ 
\displaystyle{\int_{0}^{\epsilon }f_{\beta _{\mathcal{F}}}\left( \beta _{\mathcal{F}%
}\right) d\beta _{\mathcal{F}}}, & \text{Otherwise}%
\end{array}%
\right. 
\end{equation}%
where $\epsilon =\sqrt{\frac{\gamma _{\mathrm{th}}\mathbb{E}\left(
Z^{2}\right) }{\alpha_{2}-4\alpha_{1}\gamma _{\mathrm{th}}}}$.
Consequently, $P_{O_{\mathcal{F}}}=1$ given that 
\begin{equation}
R>\log _{2}\left( \frac{\alpha _{2}}{4\alpha _{1}}+1\right). 
\end{equation}
Evaluating the integral in \eqref{eq:POF-3} gives,
\begin{multline}
P_{O_{\mathcal{F}}} =\frac{1}{2p_{\mathcal{F}}}\left( 1-2\mathrm{Q}{\left( 
{\frac{\left\vert X_{ij}\right\vert \epsilon }{\sigma_{\mathrm{n}}}}\right) }{%
\mathrm{e}^{-{\frac{\epsilon ^{2}}{\Omega }}}}\right) \\
-\frac{\left\vert X_{ij}\right\vert }{2p_{\mathcal{F}}\varphi _{\mathcal{F}}}{%
\mathrm{erf}\left( \epsilon \sqrt{{\frac{\left\vert X_{ij}\right\vert ^{2}\bar{%
\gamma}+2\,}{2\Omega }}}\right) }  \label{eq:POF-2}
\end{multline}%
where $\varphi _{\mathcal{F}}=\sqrt{{\frac{\left\vert X_{ij}\right\vert ^{2}\bar{%
\gamma}+2}{\bar{\gamma}}}}$. 
Finally, the \gls{op} can be computed as, 
\begin{align}
\label{eq:PO_total}
P_{O} &=\sum_{i}\sum_{j}P_{O}\left( \mathcal{S},X_{ij}\right) +P_{O}\left( \mathcal{F}%
,X_{ij}\right)   \nonumber \\
&=\Pr \left( X_{ij}\right) \sum_{i}\sum_{j}P_{O_{\mathcal{S}}}\text{ }p_{%
\mathcal{S}}+P_{O_{\mathcal{F}}}p_{\mathcal{F}}
\end{align}
where $P_{O}\left(\mathcal{S/F},X_{ij}\right)$ is the \gls{op} given a successful/failure \gls{sic} with $X_{ij}$ being transmitted, $p_{\mathcal{S}}$ is given in \eqref{eq:ps-1} and $p_{\mathcal{F}}=1-p_{\mathcal{S}}$, and the symbols $X_{ij}$ are equiprobable. As shown in \eqref{eq:PO_total}, the exact \gls{op} is expressed in closed-form and is parameter-free.

\color{black}

\section{Ergodic Capacity Analysis}
\label{sec: EC_Analysis}

Following the outage analysis approach, the \gls{ec} of $U_2$ can be expressed as
\begin{align}
\bar{C}&=\bar{C}_{\mathcal{S}}p_{\mathcal{S}}+\bar{C}_{\mathcal{F}}p_{\mathcal{F}} \nonumber\\
    & = \mathbb{E}\big[\log_2(1+\gamma_{\mathcal{S}})\big]p_{\mathcal{S}}
+\mathbb{E}\big[\log_2(1+\gamma_{\mathcal{F}})\big]p_{\mathcal{F}}.
    \label{eq:C2_def}
\end{align}
It is worth noting that all terms in \eqref{eq:C2_def} are conditioned on $X_{i,j}$. To compute the expectation with respect to $\beta_{\mathcal{S}}$ and $\beta
_{\mathcal{F}}$,
\begin{equation}
\label{eq:C-00}
\!\!\!    \bar{C} = \int_0^{\infty} p_{\mathcal{S}} C_{\mathcal{S}} f_{\beta_{\mathcal{S}}}(\beta_{\mathcal{S}}) \, d\beta_{\mathcal{S}} + \int_0^{\infty} p_{\mathcal{F}} C_{\mathcal{F}} f_{\beta_{\mathcal{F}}}(\beta_{\mathcal{F}}) \, d\beta_{\mathcal{F}}
\end{equation}
where $C_{\mathcal{S}/\mathcal{F}}=\log_2(1+\gamma_{\mathcal{S}/\mathcal{F}})$. By substituting the expressions of $\gamma_{\mathcal{S}}$ and $f_{\beta_{\mathcal{S}}}(\beta_{\mathcal{S}})$, and noting that $\Phi(x)=1-Q(x)$, then
\begin{align}
\label{eq:CS-00}
\bar{C}_{\mathcal{S}} &=\int_{0}^{\infty }\log _{2}\left( 1+\frac{\alpha
_{2}\beta _{\mathcal{S}}^{2}}{\mathbb{E}[W^{2}]}\right) f_{\beta_{\mathcal{S%
}}}(\beta_{\mathcal{S}})d\beta _{\mathcal{S}} \nonumber\\
&=\frac{2}{\Omega p_{\mathcal{S}}}\int_{0}^{\infty }\beta_{\mathcal{S}%
}\log_{2}\left( 1+\frac{\alpha _{2}\beta _{\mathcal{S}}^{2}}{\mathbb{E}%
[W^{2}]}\right) \nonumber\\
& \hspace{2 cm} \times\left[ 1-Q\left( \frac{|X_{ij}|\beta _{\mathcal{S}}}{\sigma
_{\mathrm{n}}}\right) \right] \mathrm{e}^{-\frac{\beta _{\mathcal{S}}^{2}}{%
\Omega }}d\beta _{\mathcal{S}}.
\end{align}
Using the distributive property of integration, the first integral in \eqref{eq:CS-00} can be written as
\begin{equation}
I_{1}=\int_{0}^{\infty }\beta _{S}\log _{2}\left( 1+\frac{\alpha _{2}\beta
_{S}^{2}}{\mathbb{E}[W^{2}]}\right) \mathrm{e}^{-\frac{\beta_{S}^{2}}{\Omega }%
}\,d\beta_{S}.
\end{equation}
Substituting $t=\beta _{S}^{2}$ and $\beta _{S}\,d\beta _{S}=\frac{dt}{2}$,
\begin{equation}
    I_{1}=\frac{1}{2}\int_{0}^{\infty }\log _{2}\left( 1+\frac{\alpha
_{2}t}{\mathbb{E}[W^{2}]}\right) \mathrm{e}^{-\frac{t}{\Omega }}\,dt.
\end{equation}
which can be evaluated using \cite[4.337-2]{gradshteyn2007table} as
\begin{equation}
    I_{1}=\frac{\Omega }{2\ln 2}\,\mathrm{e}^{\frac{\mathbb{E}[W^{2}]}{\alpha _{2}\Omega }%
}\,E_{1}\!\left( \frac{\mathbb{E}[W^{2}]}{\alpha _{2}\Omega }\right) . 
\end{equation}

The second integral in \eqref{eq:CS-00} includes the $Q$-function, and hence does not have a closed-form solution. Therefore, it can be evaluated efficiently using the Gauss–Laguerre or adaptive quadrature methods. Combining the two expressions gives
\begin{multline}
\bar{C}_{\mathcal{S}}=\frac{1}{p_{\mathcal{S}}\ln 2}\,\mathrm{e}^{\frac{%
\mathbb{E}[W^{2}]}{\alpha _{2}\Omega }}\,E_{1}\left( \frac{\mathbb{E}[W^{2}]}{\alpha
_{2}\Omega }\right) -\frac{2}{\Omega p_{\mathcal{S}}}\int_{0}^{\infty }\beta
_{\mathcal{S}}\\
\times\log_{2}\left( 1+\frac{\alpha _{2}\beta_{\mathcal{S}}^{2}}{%
\mathbb{E}[W^{2}]}\right) Q\left( \frac{|X_{ij}|\beta _{\mathcal{S}}}{\sigma
_{\mathrm{n}}}\right) \mathrm{e}^{-\frac{\beta _{\mathcal{S}}^{2}}{\Omega }%
}d\beta _{\mathcal{S}}.
\end{multline}

The second component of the \gls{ec} corresponds to $C_{\mathcal{F}}$, which has the same form as the second part of \eqref{eq:CS-00} that includes the $Q$-function, in addition to the more involved  $\gamma_{\mathcal{F}}$ expression. Therefore, the integral in this part can be evaluated numerically as well. The average overall capacity 
\begin{equation}
\bar{C}_A=\Pr(X_{ij})\sum_i\sum_j\bar{C}_{\mathcal{S}}p_{\mathcal{S}}+\bar{C}_{\mathcal{F}}p_{\mathcal{F}}.
    \label{eq:CA-00}
\end{equation}
Although \eqref{eq:CA-00} is exact, it requires numerical integration. Therefore, deriving an accurate approximation would be very beneficial. Toward this goal, it should be noted that at $p_{\mathcal{F}} \bar{C}_{\mathcal{F}}\ll p_{\mathcal{S}}\bar{C}_{\mathcal{S}}$ because at high \glspl{snr} $p_{\mathcal{F}} \ll p_{\mathcal{S}}$ and at low \gls{snr} $\bar{C}_{\mathcal{F}}\ll \bar{C}_{\mathcal{S}}$. Therefore  \eqref{eq:CA-00} can be approximated by
\begin{equation}
\bar{C}_A\approx\Pr(X_{ij})\sum_i\sum_j\bar{C}_{\mathcal{S}}p_{\mathcal{S}}.
    \label{eq:CA-App-00}
\end{equation}
Moreover, applying the Chiani approximation for the Q-function \cite{chiani2003new}, $Q(x) \approx \frac{1}{%
12} \mathrm{e}^{-x^2/2} + \frac{1}{4} \mathrm{e}^{-2x^2/3}$, 
the second integral in \eqref{eq:CS-00} can be written as
\begin{multline}
I_2 \approx 
\int_{0}^{\infty} \beta_{\mathcal{S}} \log_2\left(1+\frac{\alpha_2 \beta_{%
\mathcal{S}}^2}{\mathbb{E}[W^2]}\right) \\
 \times \left[ \frac{1}{12} \mathrm{e}^{-\left(\frac{|X_{ij}|^2}{2\sigma_{%
\mathrm{n}}^2} + \frac{1}{\Omega}\right)\beta_{\mathcal{S}}^2} + \frac{1}{4}
\mathrm{e}^{-\left(\frac{2|X_{ij}|^2}{3\sigma_{\mathrm{n}}^2} + \frac{1}{\Omega}%
\right)\beta_{\mathcal{S}}^2} \right] d\beta_{\mathcal{S}}.
\end{multline}
By defining the general integral form $I(A, B) = \int_{0}^{\infty} \beta \log_2(1
+ A \beta^2) \mathrm{e}^{-B \beta^2} d\beta$, where $A = \frac{\alpha_2}{\mathbb{E}%
[W^2]}$, $B_1 = \frac{|X_{ij}|^2}{2\sigma_{\mathrm{n}}^2} + \frac{1}{\Omega}$%
, and $B_2 = \frac{2|X_{ij}|^2}{3\sigma_{\mathrm{n}}^2} + \frac{1}{\Omega}$, 
and using identity \cite[4.337-2]{gradshteyn2007table}, we obtain 
\begin{equation}
I_2 \approx \frac{1}{4} \left[ 
\frac{1}{3} I(A, B_1) + I(A, B_2) \right]
\end{equation}
where 
\begin{equation}
 I(A, B) = \frac{1}{2 \ln(2) B} \mathrm{e}^{B/A} E_1\left(\frac{B}{A}\right)
\end{equation}
Expanding the result with the parameters $A,B_{1},B_{2}$: 
\begin{equation}
\hspace{-4mm}I_2\approx \frac{1}{8\ln (2)}\left[
\frac{1}{3B_{1}}\mathrm{e}^{\frac{B_{1}}{A}}E_{1}\left( \frac{B_{1}}{A}%
\right) +\frac{1}{B_{2}}\mathrm{e}^{\frac{B_{2}}{A}}E_{1}\left( \frac{B_{2}}{A}%
\right) \right]. 
\end{equation}
Finally, 
\begin{equation}
\label{eq:CA-01}
    \bar{C}_A\approx\frac{2}{\Omega}\Pr(X_{i,j}) \sum_i\sum_j I_1+I_2.
\end{equation}
Table \ref{tab:App_vs_Lega} compares the accuracy of the derived approximation versus the legacy formula in \eqref{eq:C_lgacy}. The comparison is performed in terms of the normalized error 
\begin{equation}
    error=\frac{|C_A(Exact)-C_A(Appr.)|}{C_A(Exact)}\times 100\%.
\end{equation}
The error for the legacy \gls{ec} is computed similarly while using $C_A(Legacy)$ instead of $C_A(Appr.)$. The error is generated for $\bar{\gamma=[0,1,\dots,30]}$ and then the minimum, maximum and average error values are recorded. As can be noted from the table, the derived approximation is significantly more accurate than the legacy formula, with a maximum average error of $1.11\%$, while the maximum average error for the legacy formula is $10.87\%$ at $\alpha_1=0.55$. The maximum error is $3.42\%$ and $17.25\%$ for the approximation and legacy formulae, respectively. Therefore, the derived closed-form capacity expression can be used with a negligible error. Similar to \gls{op}, the \gls{ec} expression is given in closed-form and is parameter-free. The exact expression has a single integral and is parameter-free. 
\begin{table*}[t]
\centering
\caption{Normalized error comparison of the approximated capacity\eqref{eq:CA-01} versus the legacy formula in \eqref{eq:C_lgacy}, $\bar{\gamma}=[0,1,\dots,30]$ {dB} and $\zeta=0$.}
\label{tab:App_vs_Lega}
\begin{tabular}{llcccccccc}
\toprule
 & & \multicolumn{8}{c}{$\alpha_1$} \\
\cmidrule(lr){3-10}
Method & Metric & 0.55 & 0.6 & 0.65 & 0.7 & 0.75 & 0.8 & 0.85 & 0.9 \\
\midrule
\multirow{3}{*}{Appr.} & Min  & 0.04\% & 0.04\% & 0.04\% & 0.02\% & 0.01\% & 0.00\% & 0.00\% & 0.00\% \\
                       & Max  & 2.34\% & 2.47\% & 2.71\% & 2.96\% & 3.17\% & 3.33\% & 3.41\% & 3.42\% \\
                       & Avg & 1.11\% & 0.57\% & 0.79\% & 0.87\% & 0.86\% & 0.81\% & 0.73\% & 0.64\% \\
\midrule
\multirow{3}{*}{Legacy} & Min & 1.52\% & 0.16\% & 0.00\% & 0.00\% & 0.04\% & 0.04\% & 0.03\% & 0.03\% \\
                        & Max & 17.25\% & 10.66\% & 5.71\% & 1.61\% & 2.48\% & 5.29\% & 8.23\% & 11.06\% \\
                        & Avg & 10.87\% & 5.24\% & 2.17\% & 0.60\% & 1.25\% & 2.33\% & 3.20\% & 3.95\% \\
\bottomrule
\end{tabular}
\end{table*}

\section{Post-SIC Noise and Fading PDFs with QPSK}\label{sec:QPSK}

Deriving the \gls{op} and \gls{ec} for \gls{qpsk} can be performed using the same approach as the \gls{bpsk}. Nevertheless, the analysis is generally more tedious because the success and failure events become two-dimensional. Therefore, this section outlines the derivation of $\gamma_{\mathcal{S}}$ while the $\gamma_{\mathcal{F}}$ can be derived using the same approach.     


\subsection{System Model with QPSK}

With \gls{qpsk}, the \gls{noma} transmitted signal $X$ is given in %
\eqref{eq:X}, but~ $\{s_1,s_2\}\in \{ \frac{1}{\sqrt{2}}(\pm 1 \pm \mathfrak{%
j})\}$. Consequently, the $16$ constellation points are given by,
\begin{equation}
\left\{ \frac{m_1 \sqrt{ \alpha_1} + m_2 \sqrt{\alpha_2}}{\sqrt{2}} ,~ 
\frac{m_3 \sqrt{ \alpha_1} + m_4 \sqrt{\alpha_2}}{\sqrt{2}} \right\}
\end{equation}
where $m_i\in\{ 1,-1\}$. Due to the symmetry of the performance with respect to the constellation point in each quadrant, without loss of generality, we
consider the top-right quadrant, which has $s_1 = \frac{1+\mathfrak{j}}{%
\sqrt{2}}$, and thus 
\begin{equation}
~ ~ \left\{ \frac{\sqrt{ \alpha_1} + i \sqrt{\alpha_2}}{\sqrt{2}} ,~ \frac{ 
\sqrt{ \alpha_1} + j \sqrt{\alpha_2}}{\sqrt{2}} \right\} \triangleq
\{\lambda_{i}, \lambda_j \}
\end{equation}
where $\{i,j \}\in\{ 1,-1\}$. The received signal at $U_2$ after phase
equalization is similar to \eqref{eq:Y}. After dropping the user index, the
unconditional joint \gls{pdf} of the Rayleigh channel gain $\beta$ and the
complex Gaussian noise $N$ is given by, 
\begin{equation}
f_{\beta ,N}(\beta ,n)=\frac{2\beta }{2\pi \Omega \sigma _{n}^{2}}\exp
\left( -\frac{\beta ^{2}}{\Omega }-\frac{|n|^{2}}{2\sigma _{n}^{2}}\right)~
\end{equation}
where $\beta \ge 0$ and~ ~ ~ ~$N = N_R + \mathfrak{j}N_I$.~

\subsection{Fading and Noise joint PDF, $f_{\protect\beta_{\mathcal{S}}, W}(%
\protect\beta_{\mathcal{S}}, w )$}

Given that $\hat{s}_{1}=s_{1}$, then the received signal $Y$ is within the
decision region $\mathcal{D}_{\mathcal{S}}$ for $s_{1}=\frac{1+\mathfrak{j}}{%
\sqrt{2}}$.~ For \gls{qpsk}, $\mathcal{D}_{\mathcal{S}}$ is given by~ 
\begin{equation}
\mathcal{D}_{\mathcal{S}}=\left\{ \Re(Y)>0\text{ , }\Im%
(Y)>0\right\} .
\end{equation}%
The condition $(\beta ,n)\in \mathcal{D}_{\mathcal{S}}$ simplifies to: $%
N_{R}>-\beta \lambda _{i}$ and $N_{I}>-\beta \lambda _{j}$. 
Using Bayes' rule for continuous variables, the conditional joint \gls{pdf}
is the original \gls{pdf}, truncated by the indicator function $\mathbb{I}%
(\cdot )$ of the decision region, and normalized by the probability of
success $p_{\mathcal{S}}$. Therefore, 
\begin{equation}
f_{\beta _{\mathcal{S}},W}(\beta _{\mathcal{S}},w)=\frac{%
f_{\beta }(\beta _{\mathcal{S}})f_{N}(w)}{p_{\mathcal{S}}}~\mathbb{I}\big(%
(\beta _{\mathcal{S}},w)\in \mathcal{D}_{\mathcal{S}}\big)
\end{equation}%
where $p_{\mathcal{S}}=\mathbb{E}_{\beta }[p_{\mathcal{S}}|\beta ]$, the
indicator function $\mathbb{I}\big((\beta _{\mathcal{S}},w)\in \mathcal{D}_{%
\mathcal{S}}\big)$ implies that for a given $w$, the success condition is
satisfied only if $\beta >\max (0,-w_{R}/\lambda _{i},-w_{I}/\lambda _{j})$,
and 
\begin{align}
\label{eq:ps_beta}
p_{\mathcal{S}}|\beta & =\left[ 1-Q\left( \beta \chi _{i}\right) %
\right] \left[ 1\!-\!Q\left( \beta \chi _{j}\right) \right]  \nonumber
 \\
& =1\!-\!Q\!\left( \beta \chi _{i}\right) \!-\!Q\!\left( \beta \chi
_{j}\right) \!+\!Q\!\left( \beta \chi _{i}\right) Q\!\left( \beta \chi
_{j}\right) 
\end{align}%
where $\chi_{i/j}=\lambda _{i/j}/\sigma _{\mathrm{n}}$, and thus, $p_{\mathcal{S%
}}=\int_{0}^{\infty }p_{\mathcal{S}}|\beta ~f_{\beta }(\beta )d\beta $.~ The
integral of the first term in \eqref{eq:ps_beta} is equal to $1$, the second
and third terms can be evaluated using the identity~ ~$\mathbb{E}_{\beta }%
\left[ Q(\beta \chi )\right] =\frac{1}{2}\left( 1-\mu \right) $, where $\mu =%
\sqrt{\frac{\lambda ^{2}\,\bar{\gamma}}{2+\lambda ^{2}\,\bar{\gamma}}}$. The
integral of the fourth term involves the product of two $Q$-functions. For $%
\lambda _{i}=\lambda _{j}$, the integral can be evaluated using the standard
identity for the average of the squared $Q$-function in Rayleigh fading,
derived from the Craig's formula representation, $\mathbb{E}_{\beta }\left[
Q^{2}(\beta \chi _{i})\right] =\frac{1}{4}-\frac{1}{\pi }\arctan (\mu _{i})$. Substituting the results and simplifying the expression yields, 
\begin{equation}
~~p_{\mathcal{S}}|[\lambda _{i}=\lambda _{j}]=~p_{\mathcal{S}}|\lambda
_{i}=\mu _{i}+\frac{1}{4}-\frac{1}{\pi }\arctan (\mu _{i}).
\label{eq:pS_i_eq_j}
\end{equation}%
For $\lambda _{i}\neq \lambda _{j}$, the integral of the fourth term does
not have a closed-form analytic solution \cite{Zainab_MR}.~ Consequently, by
noting the geometry of the four constellation points, and noting that the first three dominant terms are given in closed-form, an accurate
approximation can be obtained as~ 
\begin{equation}
~~~p_{\mathcal{S}}|[\lambda _{i}\neq \lambda _{j}]=\frac{1}{2}p_{\mathcal{S}%
}|\lambda _{i}+\frac{1}{2}p_{\mathcal{S}}|\lambda _{j}.
\label{eq:pS_i_neq_j}
\end{equation}%
It can be noted that \eqref{eq:pS_i_eq_j} is a special case of %
\eqref{eq:pS_i_neq_j}, and thus, both can be generally expressed as~ 
\begin{equation}
~~~p_{\mathcal{S}}=\frac{1}{2}p_{\mathcal{S}}|\lambda _{i}+\frac{1}{2}p_{%
\mathcal{S}}|\lambda _{j}.  \label{eq:pS_general}
\end{equation}%
It is worth noting that \eqref{eq:pS_general} is exact when $\lambda
_{i}=\lambda _{j}$ and near exact $\lambda _{i}\neq \lambda _{j}$, with a
normalized error that is less than $1\%$ for high \glspl{snr}.

\subsection{Marginal PDF of $\protect\beta_{\mathcal{S}}$}

The marginal \gls{pdf} of the fading coefficient $f_{\beta _{\mathcal{S}%
}}(\beta _{\mathcal{S}})$ can be computed using the joint \gls{pdf} $%
f_{\beta _{\mathcal{S}},W}(\beta _{\mathcal{S}},w)$. However, to avoid the
double integral, we use Bayes' rule, 
\begin{equation}
~~f_{\beta _{\mathcal{S}}}(\beta _{\mathcal{S}})=\frac{1}{p_{\mathcal{S}}}%
~f_{\beta }(\beta _{\mathcal{S}})~~p_{\mathcal{S}}|\beta _{\mathcal{S}}.
\end{equation}%
By substituting the corresponding terms, 
\begin{multline}
~~f_{\beta _{\mathcal{S}}}(\beta _{\mathcal{S}})=~~\frac{2\beta _{\mathcal{S}%
}}{\Omega p_{\mathcal{S}}}\exp \left( -\frac{\beta _{\mathcal{S}}^{2}}{%
\Omega }\right) \left[ 1-Q\left( \beta _{\mathcal{S}}\chi _{i}\right) \right]
\\
~~\times \left[ 1-Q\left( \beta _{\mathcal{S}}\chi _{j}\right) \right] .
\end{multline}

\subsection{Marginal PDF of Conditional Noise $f_{W}(w)$}

The marginal \gls{pdf} of the complex noise $W$ is obtained by integrating
the joint conditional \gls{pdf} over $\beta_{\mathcal{S}}$, 
\begin{equation}
~ ~ f_{W}(w) = \int_{0}^{\infty} \frac{f_{\beta}(\beta_{\mathcal{S}})
f_{N}(w)}{p_{\mathcal{S}}} \mathbb{I}\big((\beta_{\mathcal{S}}, w) \in 
\mathcal{D}_{\mathcal{S}}\big) \, d\beta_{\mathcal{S}}.
\end{equation}
~By denoting $\tau(w) = \max(0, -w_R/\lambda_i, -w_I/\lambda_j)$, the
integral becomes 
\begin{equation}  \label{eq:f_W_qpsk}
~ ~ f_{W}(w) = \frac{f_{N}(w)}{p_{\mathcal{S}}} \int_{\tau(w)}^{\infty} 
\frac{2\beta_{\mathcal{S}}}{\Omega} \exp\left( -\frac{\beta_{\mathcal{S}}^2}{%
\Omega} \right) d\beta_{\mathcal{S}}.
\end{equation}
Evaluating the integral yields, 
\begin{equation}
 \!\!\!\!f_{W}(w)\! =\! \frac{f_{N}(w)}{p_{\mathcal{S}}} \exp\!\left( -%
\frac{[\max(0, -w_R/\lambda_i, -w_I/\lambda_j)]^2}{\Omega} \right)
\end{equation}
where $f_{N}(w) = \frac{1}{2\pi \sigma_n^2} \exp\left( -\frac{|w|^2}{%
2\sigma_n^2} \right)$ is the unconditional \gls{awgn} \gls{pdf}. It is worth
noting that, in contrast to \gls{bpsk}, the real and imaginary components of 
$W$ became dependent after being conditioned on the \gls{sic} outcome.

\subsection{Marginal PDF of the Real the Noise $f_{W_R}(w_R)$}

The conditional marginal \gls{pdf} of the real noise component $W_{R}$ is
obtained by averaging the joint conditional \gls{pdf} \eqref{eq:f_W_qpsk}, 
\begin{multline}
f_{W_R}(w_{R})=\frac{f_{N_{R}}(w_{R})}{p_{\mathcal{S}}}\int_{-\infty
}^{\infty }f_{N_{I}}(w_{I}) \\
~~\times \exp \left( -\frac{[\max (0,-w_{R}/\lambda _{i},-w_{I}/\lambda
_{j})]^{2}}{\Omega }\right) dw_{I}.
\end{multline}%
The integral can be evaluated piecewise based on the sign of $w_{R}$. For $%
w_{R}\geq 0$ 
\begin{equation}
f_{W_R}(w_{R})=\frac{f_{N_{R}}(w_{R})}{2p_{\mathcal{S}}}(1+\mu _{j})
\label{eq:f_W_R_QPSK_1}
\end{equation}%
and for $w_{R}<0$%
\begin{multline}
f_{W_{R}}(w_{R})=\frac{f_{N_{R}}(w_{R})}{p_{\mathcal{S}}}\left[ \mu _{j}\Phi
\left( \frac{\chi _{j}w_{R}}{\lambda _{i}\mu _{j}}\right) \right. ~
\label{eq:f_W_R_QPSK_2} \\
\left. +\mathrm{e}^{-\frac{w_{R}^{2}}{\lambda _{i}^{2}\Omega }}\Phi \left( -%
\frac{\chi _{j}w_{R}}{\lambda _{i}}\right) \right] .~
\end{multline}%
Due to the symmetry of the \gls{qpsk} constellation and the \gls{awgn}
distribution, the marginal \gls{pdf} for the imaginary part $w_{I}$ is
identical to the real part.

\subsection{Derivation of $\mathbb{E}[|W|^2]$}

Based on the similarity of \glspl{pdf} $f_{W_R}(w_R)$ and $f_{W_I}(w_I)$,
then $\mathbb{E}[|W|^2]=2\mathbb{E}[W_{R}^2]$, which is derived in the
following theorem

\begin{theorem}
\label{theorem:sec_mom} The second moment of the complex noise, given that $%
\hat{s}_1=s_1$ is given by 
\begin{align}
\mathbb{E}[|W|^{2}] &=2\mathbb{E}[W_{R}^{2}]~  \nonumber \\
&=\frac{2}{p_{\mathcal{S}}}\left[ \frac{\sigma _{\mathrm{n}%
}^{2}(1+\mu _{j})}{4}+\mu _{j}\Psi (\sigma _{\mathrm{n}},\varpi _{1})+\mu
_{i}\Psi (\sigma _{\mathrm{n}}\mu _{i},\varpi _{2})\right] 
\end{align}
where $\varpi_{1}=\frac{\chi _{j}}{\lambda_{i}\mu _{j}}$, $\varpi _{2}=-\frac{\chi _{j}}{\lambda _{i}}$, and the kernel $$%
\Psi (a,b)=\frac{a^{2}}{2}\left[ \frac{1}{2}-\frac{ba}{\sqrt{2\pi
(1+b^{2}a^{2})}}\right]$$

\begin{IEEEproof}
 The proof of $\mathbb{E}[W_{R}^{2}]$ is given in Appendix \ref{appndx-B}.
\end{IEEEproof}
\end{theorem}

\begin{table}[t]
\centering
\caption{Comparison of the variance and second moment of $W$, $\mathbb{E}[\cdot]$ is the theory and  $\tilde{\mathbb{E}}[\cdot]$ is the simulation for different SNR values.}
\label{tab:Ew2}
\renewcommand{\arraystretch}{0.9}
\begin{tabular}{m{0.5cm}m{1.9cm}m{1.0cm}m{1.0cm}m{1.0cm}m{1.0cm}}
\toprule SNR & Metric & $\lambda_{1},\lambda _{1}$ & $\lambda
_{-1},\lambda_{-1}$ & $\lambda _{1},\lambda_{-1}$ & $\lambda _{-1},\lambda
_{1}$ \\ 

\midrule\multirow{3}{*}{0} & $2\sigma_{\mathrm{n}}^{2}$ & $2$ & $2$ & $2$ & $2$ \\ 
& var$\left[ W\right]\times 10^{-1} $ & $3.69$ & $4.50$ & $4.09$ & $4.09$ \\ 
& $\mathbb{E}[|W|^{2}]$ & $1.53$ & $1.64$ & $1.41$ & $1.65$ \\ 
& $\mathbb{\tilde{E}}[|W|^{2}]$ & $1.51$ & $1.70$ & $1.60$ & $1.60$ \\ 

\midrule\multirow{3}{*}{10} & $2\sigma_{\mathrm{n}}^{2}$ & $0.2$ & $0.2$ & $%
0.2$ & $0.2$ \\ 
& var$\left[ W\right] \times 10^{-2}$ & $3.98$ & $3.79$ & $3.92$ & $3.92$ \\ 
& $\mathbb{E}[|W|^{2}]\times 10^{-1}$ & $1.86$ & $1.50$ & $1.77$ & $1.52$ \\ 
& $\mathbb{\tilde{E}}[|W|^{2}]\times 10^{-1}$ & $1.84$ & $1.49$ & $1.67$ & $%
1.67$ \\ 

\midrule\multirow{3}{*}{20} & $2\sigma_{\mathrm{n}}^{2}$ & $0.02$ & $0.02$ & 
$0.02$ & $0.02$ \\ 
& var$\left[ W\right] \times 10^{-3}$ & $4.25$ & $3.89$ & $4.07$ & $4.07$ \\ 
& $\mathbb{E}[|W|^{2}]\times 10^{-2}$ & $1.98$ & $1.81$ & $1.97$ & $1.80$ \\ 
& $\mathbb{\tilde{E}}[|W|^{2}]\times 10^{-2}$ & $1.98$ & $1.80$ & $1.89$ & $%
1.89$ \\ 
\bottomrule &  &  &  &  & 
\end{tabular}
\end{table}

 By deriving $f_{\beta_{\mathcal{S}}}(\beta_{\mathcal{S}})$ and $\mathbb{E}[|W|^{2}]$ then $P_{O_\mathcal{S}}$ can be computed by following the same steps in \eqref{PO_S_BPSK_1}. 

Table \ref{tab:Ew2} compare the analytical and simulated values of $\mathbb{E}(|W|^2)$. The table also presents the variance of $N$ and $W$. As the table shows, the analytical and simulation results match very well. The impact of the conditioning process is manifested as a lower variance than the \gls{awgn}. It can also be seen that $\mathbb{E}(|W|^2) > var[W]$ because the noise after a successful \gls{sic} has a nonzero mean.
 


\color{black}
\section{Numerical Results}\label{sec:num}
This section presents the numerical results for \gls{op} and \gls{ec} using the derived exact formulae and compares them with those obtained using the legacy expressions. The results are presented for various operating conditions. 
Monte Carlo simulation results are generated to
corroborate the accuracy of the derived analytical expressions using $\Omega=1$. Each simulation point is generated using $10^7$ independent simulation runs. All results are presented for the \gls{nu}, i.e., $U_2$.
 
\begin{figure}[t]
  \centering
  \includegraphics[width=\linewidth]{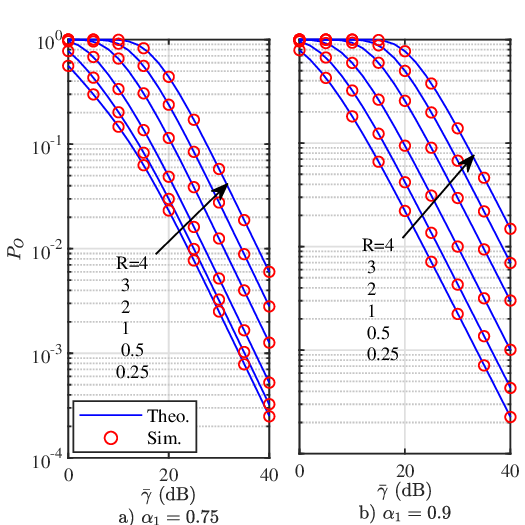} 
  \caption{\Gls{op} of $U_2$ with various target rates, $\alpha_1\in \{0.75, 0.9\}$. }
  \label{fig:Po_vs_snr}
\end{figure}

\begin{figure}[t]
  \centering
  \includegraphics[width=\linewidth]{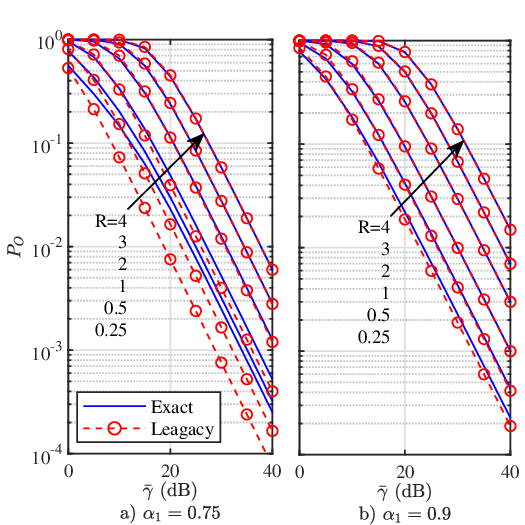} 
  \caption{$U_2$ Exact and legacy \gls{op} \eqref{eq:PO_legacy} for different target rates and power coefficients, $\zeta=0$.   }
  \label{fig:Po_vs_snr_leg}
\end{figure}

\begin{figure}[t]
  \centering
  \includegraphics[width=\linewidth]{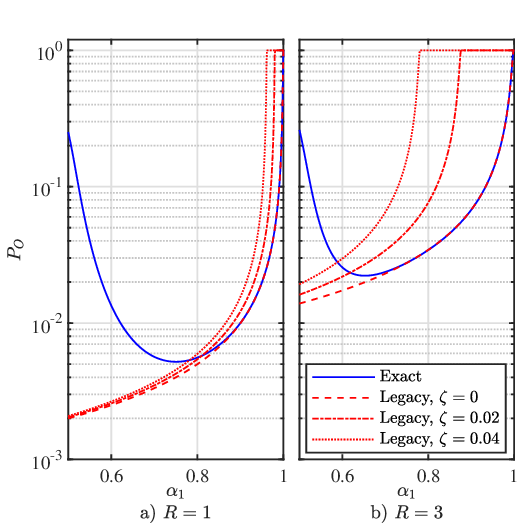} 
  \caption{Exact and legacy \gls{op} \eqref{eq:PO_legacy} versus $\alpha_1$ for various values of $\zeta$ and target rates, $R=[1,3]$, $\bar{\gamma}=30$ dB.   }
  \label{fig:Po_vs_alpha}
\end{figure}

\begin{figure}[t]
  \centering
  \includegraphics[width=\linewidth]{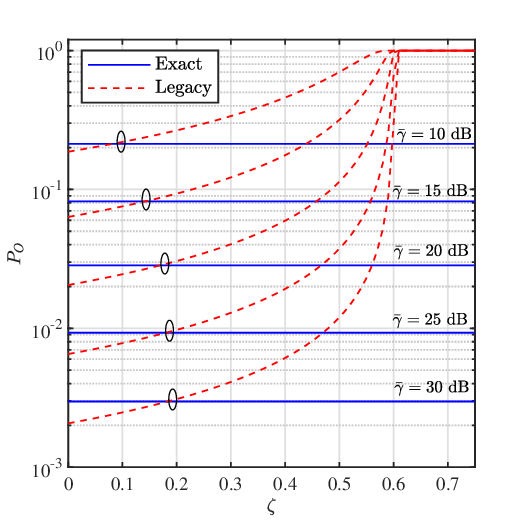} 
  \caption{$U_2$ Exact and legacy \gls{op} versus $\zeta$ for various values of $\bar{\gamma}$, $R=0.5$ and $\alpha_1=0.8$.   }
  \label{fig:Po_vs_zeta_8}
\end{figure}

\begin{figure}[t]
  \centering
  \includegraphics[width=\linewidth]{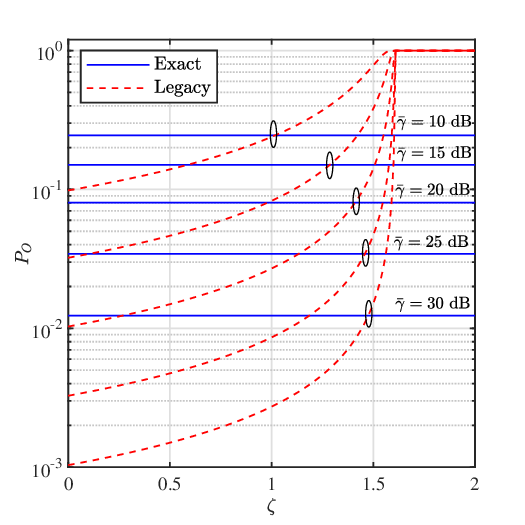} 
  \caption{$U_2$ Exact and legacy \gls{op} \eqref{eq:PO_legacy} versus $\zeta$ for various values of $\bar{\gamma}$, $R=0.5$ and $\alpha_1=0.6$.}
  \label{fig:Po_vs_zeta_6}
\end{figure}

Fig. \ref{fig:Po_vs_snr} compares the exact analytical and simulated \gls{op} results versus \gls{snr} for $R \in \{0.25, 0.5, 1, 2, 3,4\}$~bps/Hz under a fixed power allocation $\alpha_1 = 0.75$ in Fig. \ref{fig:Po_vs_snr}a and $\alpha_1 = 0.9$ in Fig. \ref{fig:Po_vs_snr}b. As shown in the figure, the analytical and simulation results match perfectly across all cases, confirming the accuracy of the analysis. Therefore, in the following figures, only the analytical results are presented to avoid overcrowding.

Fig. \ref{fig:Po_vs_snr_leg} is similar to Fig. \ref{fig:Po_vs_snr} except that the simulation results are replaced with the legacy \gls{op} in \eqref{eq:PO_legacy} using $\zeta=0$. As shown in Fig. \ref{fig:Po_vs_snr}a, the legacy \gls{op} is noticeably less than the exact for $R\leq 1$. For $R=[2,3,4]$, the exact and legacy converge and become nearly equal. Using $\zeta=0$ implies that the $P_O(\text{Legacy})\!\!<\!\! P_O(\text{Exact})$ because the impact of the residual interference is neglected. Moreover, the legacy \gls{op}  is very sensitive to the values of $R$ and $\alpha_2$. For low values of $R$ and high values of $\alpha_2$, the exponent term in \eqref{eq:PO_legacy} approaches $1$, leading to a significant drop in \gls{op}, which is not the case the exact outage because the impact of the residual interference is considered, the \gls{op} behavior is more stable. In Fig. \ref{fig:Po_vs_snr_leg}b, increasing $\alpha_1$ to $0.9$ increases \gls{op} significantly compared to $\alpha_2$ in Fig. \ref{fig:Po_vs_snr_leg}a. For high values of $\alpha_1$, the probability of \gls{sic} error decreases, and thus the system behavior approaches the one with perfect \gls{sic}. Therefore, the legacy and exact \gls{op} converge even for low rates. A key insight from this figure is that using $\zeta>0$ with high rates results in a large deviation from the exact \gls{op}. Consequently, adopting the perfect \gls{sic} model in such cases makes the legacy results close to the exact ones.

Fig. \ref{fig:Po_vs_alpha} shows the \gls{op} versus $\alpha_1$ for $R =[1,3]$ bps/Hz. In \gls{noma}, increasing $\alpha_{1}$ assigns less power to $U_2$, $\alpha_2 = 1-\alpha_1$. Therefore, increasing $\alpha_1$ will have two counteracting effects. On one hand, increasing $\alpha_1$ improves the \gls{sic} success probability, which reduces the residual interference and decreases \gls{op}. On the other hand, increasing $\alpha_1$ decreases $\alpha_2$, and thus reduces $U_2$ signal power, and increases $U_2$ \gls{op}. Consequently,  the exact outage has a concave-up shape with a minimum that depends on the rate $R$. For $R=1$ in Fig. \ref{fig:Po_vs_alpha}a, the minimum is obtained for $\alpha_1\approx0.75$, while it is obtained at $\alpha_1\approx0.65$ for $R=3$ in Fig. \ref{fig:Po_vs_alpha}b. The legacy expression does not demonstrate the same behavior because increasing $\alpha_2$ consistently decreases the exponent term in \eqref{eq:PO_legacy}, and thus increases \gls{op}. As the figure shows, varying the value of $\zeta$ does not improve the \gls{op} gap between the exact and legacy analysis because the legacy \gls{op} expression is monotonic versus $\zeta$. Generally, for high $\alpha_1$ values, using $\zeta=0$ yields the closest approximation to the exact result. Because \gls{op} is proportional to $\zeta$, increasing $\zeta$ shifts the \gls{op} curves up, which increases the gap between the exact and the legacy results. Consequently, the value of $\zeta$ should be selected based on the values of $\alpha_1$. For example, for $R=1$ and $\alpha_1\ge 0.88$, using $\zeta=0$ provides the best match with the exact \gls{op}. For $\alpha_1\sim [0.78, 0.81]$, $\zeta=0.04$ and $0.02$, respectively. The same trends are observed for $R=3$ in Fig. \ref{fig:Po_vs_alpha}b, where $\zeta=0$ should be used for $\alpha_1\ge 0.73$. In general, it can be concluded that $\zeta$ depends on the system parameters $R$, $\alpha$, and \gls{snr}.

Figs. \ref{fig:Po_vs_zeta_8} and \ref{fig:Po_vs_zeta_6} compare the exact and legacy \gls{op} versus the imperfect \gls{sic} factor $\zeta$ using various \glspl{snr}. In both figures, it can be seen that increasing \gls{snr} increases the required values of $\zeta$, and $\zeta\in[0, \frac{\alpha_2}{\alpha_1(2^R-1)}]$. For $\alpha_1=0.8$ in Fig. \ref{fig:Po_vs_zeta_8}a, $\zeta\in[0, 6.036]$, while for $\alpha_1=0.6$ in Fig. \ref{fig:Po_vs_zeta_8}b, $\zeta\in[0, 1.6095]$. It can be noted from both figures that $\zeta$ is proportional to \gls{snr}, and the change in $\zeta$ becomes very small for high \glspl{snr}. Consequently, the figures show that the range of $\zeta$ for the legacy analysis depends on the power assignment factors $\{\alpha_1, \alpha_2\}$ and the target rate $R$. Moreover, the high-end of the range equals $\frac{\alpha_2}{\alpha_1(2^R-1)}$, which can be larger than the usually adopted limit of $\zeta=1$.

\begin{figure}[t]
  \centering
  \includegraphics[width=\linewidth]{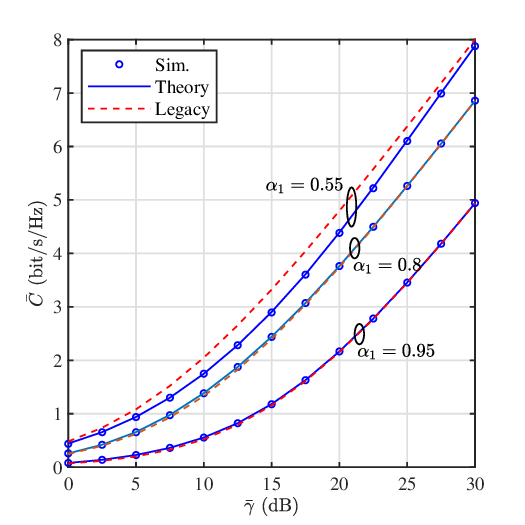}
  \caption{Exact and legacy \gls{ec} \eqref{eq:C_lgacy}versus \gls{snr} for various values of $\alpha_1$ and $\zeta=0$.}
  \label{fig:C_vs_snr_alpha}
\end{figure}

Fig.~\ref{fig:C_vs_snr_alpha} illustrates the \gls{ec} versus \gls{snr} for $\alpha_{1}\!\in\!\{0.55,0.80,0.95\}$ and compares the proposed exact post-\gls{sic} model with the legacy and Monte Carlo simulation results. The legacy results are generated using $\zeta=0$. By noting that $\gamma_{\mathcal{S}}$ and $\gamma_{\mathcal{F}}$, for a given $\{\beta_{\mathcal{S}}$ and $\beta_{\mathcal{F}} \}$, consistently increase versus $\alpha_2$, then the capacity is expected to increase by decreasing $\alpha_1$. Interestingly, the exact and legacy capacities match closely, even though $\zeta=0$, except for very small values of $\alpha_1$. Therefore, the results in the figure suggest that using $\zeta=0$ with legacy analysis provides reasonable accuracy for practical power-assignment scenarios. 

\begin{figure}[!t]
  \centering
  \includegraphics[width=\linewidth]{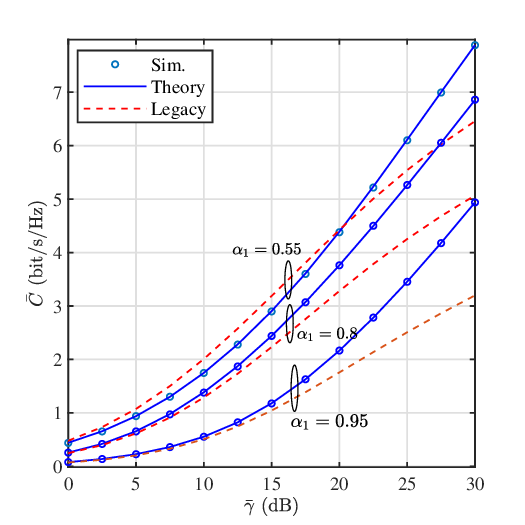}
  \caption{Exact and legacy \gls{ec} \eqref{eq:C_lgacy} versus \gls{snr} for various values of $\alpha_1$, $\zeta=0.01$.}
  \label{fig:C_vs_snr_alpha_zeta}
\end{figure}

Fig. \ref{fig:C_vs_snr_alpha_zeta} is similar to Fig.~\ref{fig:C_vs_snr_alpha} except that $\zeta=0.01$. As the figure shows, increasing $\zeta$, even by small values, significantly impacts the capacity, making it well below the exact for $\alpha_1=0.8$ and $0.95$. For $\alpha_1=0.55$, the legacy capacity is larger than the exact at low and moderate \glspl{snr}, while it becomes smaller at high \glspl{snr}. In line with Fig. \ref{fig:C_vs_snr_alpha_zeta}, the results in this figure also suggest that using $\zeta=0$ with capacity analysis provides the nearest results to the exact.

 \section{Conclusion and Future Work}
 \label{sec:conclusion}
This work developed an exact analytical framework for the \gls{op} and \gls{ec} of the \gls{nu} in a two-user downlink \gls{noma} system over Rayleigh fading with \gls{bpsk} modulation. By noting the impact of a successful and failure \gls{sic} processes on the statistical properties of the noise and fading, the joint \gls{pdf} of noise and fading is derived and used to derive the marginal \glspl{pdf} in the case of successful and failure \gls{sic}. Then, the marginal \glspl{pdf} are used to derive the exact \gls{op} and \gls{ec} and compare them to the legacy analysis, which considers that the noise and fading are independent of the \gls{sic} outcome. The main advantage of the derived expressions is that they do not require the imperfect \gls{sic} tuning factor. 
To provide deep insights into the behavior of the legacy versus exact analysis, extensive numerical results were generated for a range of \glspl{snr}, power-allocation factors, and target rates. The results showed that $\zeta$ cannot be selected arbitrarily because it depends on all system parameters, i.e., \gls{snr}, target rate, and power allocation. In the case of \gls{op}, using $\zeta=0$ provides accurate results for $R\geq 1$ and $\alpha_1 \gg 0.5$, regardless of the \gls{snr}. For the \gls{ec}, using $\zeta=0$ generally yields the most accurate results, whereas $\zeta > 0$ can significantly degrade the computed \gls{ec}.

Future work may extend the analysis to higher-order modulations, multi-antenna and multi-user \gls{noma} with various channel models. Moreover, the new exact analysis can be applied to evaluate the performance of existing works that use the \gls{sic} imperfection parameter $\zeta$ in an arbitrary manner. 

\appendices

\section{Proof of Theorem \ref{Theo:PDFs-1}}
\label{appndx-A}
 Let $\beta\sim \mathcal{R}(\Omega)$ and $N\sim \mathcal{N}(0, \sigma_{\mathrm{n}}^{2})$ be independent Rayleigh and Gaussian random variables.
 Therefore, their joint \gls{pdf} is the product of
their marginals 
\begin{equation}
f_{\beta ,N}(\beta ,n)=f_{\beta }(\beta )f_{N}(n)
\end{equation}
Given that $Y=\beta X_{ij}+N$, $X_{ij}\in \left\{ X_{11},X_{10}\right\}>0 $, then the necessary and sufficient condition for a successful \gls{sic} is that $Y\geq 0$. 
The condition $Y\geq 0$ implies that the joint \gls{pdf} $f_{\beta _{\mathcal{S}%
},W}(\beta _{\mathcal{S}},w)$ is a truncated version of $f_{\beta ,N}(\beta ,n),
$ where $Y\geq 0$ corresponds to the region $n\geq -X_{ij}\beta $.
Therefore, 
\begin{equation}
f_{\beta _{\mathcal{S}},W}(\beta _{\mathcal{S}},w)=\frac{f_{\beta ,N}(\beta
,n)}{p_{\mathcal{S}}}\text{, \ }n\geq -X_{ij}\beta 
\end{equation}
where $p_{\mathcal{S}}=\Pr(Y\geq 0)$, 
\begin{align}
p_{\mathcal{S}} &=\int_{0}^{\infty }f_{\beta }(\beta )\left[
\int_{-X_{ij}\beta }^{\infty }f_{N}(n)\,dn\right] d\beta  \nonumber \\
&=\int_{0}^{\infty }f_{\beta }(\beta )\Phi \left( \frac{X_{ij}\beta }{%
\sigma _{\mathrm{n}}}\right) d\beta \nonumber \\
&=\frac{1}{2}+\frac{1}{2}\sqrt{\frac{X_{ij}^{2}\bar{\gamma}}{X_{ij}^{2}\bar{%
\gamma}+2}}
\end{align}%

To find the marginal PDF of $\beta _{\mathcal{S}}$, integrate over the
valid range of $n$, 
\begin{align}
f_{\beta _{\mathcal{S}}}(\beta _{\mathcal{S}}) &=\int_{-X_{ij}\beta _{%
\mathcal{F}}}^{\infty }\frac{f_{\beta }(\beta _{\mathcal{S}})f_{N}(n)}{p_{%
\mathcal{S}}}dn \nonumber\\
&=\frac{f_{\beta }(\beta _{\mathcal{S}})}{p_{\mathcal{S}}}\Phi \left( \frac{%
X_{ij}\beta_{\beta _{\mathcal{S}}}}{\sigma _{\mathrm{n}}}\right) 
\end{align}

The marginal \gls{pdf} of $N$ depends on the sign of $n$ due to the constraint $%
\beta _{\mathcal{S}}\geq \max (0,-n/X_{ij})$.

\begin{itemize}
\item {Case 1: $n\geq 0$}: 
The condition $\beta \geq -n/X_{ij}$ is always satisfied for $\beta \geq 0$. Thus,
\begin{equation}
\!\!f_{W}(w) =\frac{f_{N}(w)}{p_{\mathcal{S}}}\int_{0}^{\infty }f_{\beta
}(\beta )\,d\beta 
=\frac{f_{N}(w)}{p_{\mathcal{S}}}\text{, }w\geq 0
\end{equation}

\item {Case 2: $n<0$}: 
The variable $\beta $ is restricted to $\beta \geq -n/X_{ij}$. Thus 
\begin{align}
f_{W}(w) &=\frac{f_{N}(w)}{p_{\mathcal{S}}}\int_{-n/X_{ij}}^{\infty
}f_{\beta }(\beta )\,d\beta  \nonumber \\
&=\frac{f_{N}(w)}{p_{\mathcal{S}}}\mathrm{e}^{-\frac{w^{2}}{%
X_{ij}^{2}\Omega }}\text{,  }w<0.
\end{align}
\end{itemize}

The probability of \gls{sic} error $p_{\mathcal{F}}=\Pr(Y<0)$, is the complement of the success probability $\Pr(Y\geq 0)$, therefore, $p_{\mathcal{F}}=1-p_{%
\mathcal{S}}$. For the case of unsuccessful \gls{sic}, the necessary and sufficient condition is that $Y<0$, i.e., the joint \gls{pdf} is restricted to the region $X_{ij}\beta+N<0$, or $N<-X_{ij}\beta$. Since $\beta \geq 0$, this implies that $N$ must be negative. Therefore, 
\begin{equation}
f_{\beta _{\mathcal{F}},Z}(\beta _{\mathcal{F}},z)=\frac{f_{\beta ,N}(\beta
_{\mathcal{F}},z)}{p_{\mathcal{F}}}\text{, }\beta _{\mathcal{F}}\geq 0\text{%
, }z<-X_{ij}\beta_{\mathcal{F}}.
\end{equation}%
Integrating the joint PDF over $n$ from $-\infty $ to $-X_{ij}\beta _{%
\mathcal{F}}$ gives the marginal \gls{pdf} of $\beta_{\mathcal{F}}$,
\begin{align}
f_{\beta _{\mathcal{F}}}(\beta _{\mathcal{F}}) &=\frac{f_{B}(\beta _{%
\mathcal{F}})}{p_{\mathcal{F}}}\int_{-\infty }^{-X_{ij}\beta _{\mathcal{F}%
}}f_{N}(n)\,dn  \notag \\
&=\frac{f_{\beta }(\beta _{\mathcal{F}})}{p_{\mathcal{F}}}Q\left( \frac{%
X_{ij}\beta _{\mathcal{F}}}{\sigma _{\mathrm{n}}}\right) \text{, }\quad
\beta _{\mathcal{F}}\geq 0.
\end{align}

For a fixed negative noise $n<0$, the condition $X_{ij}\beta +N<0$ implies $%
\beta <-n/X_{ij}$. Therefore, the marginal \gls{pdf} of $Z$ can be obtained by integrating the joint \gls{pdf} over $\beta \in
\lbrack 0,-n/X_{ij}]$, 
\begin{equation}
f_{Z}(z)=\frac{f_{N}(z)}{p_{\mathcal{F}}}\int_{0}^{-n/X_{ij}}f_{\beta
}(\beta _{\mathcal{F}})\,d\beta _{\mathcal{F}}.
\end{equation}%
Evaluating the Rayleigh integral (which is the Rayleigh CDF): 
\begin{equation}
f_{Z}(z)=\frac{f_{N}(z)}{p_{\mathcal{F}}}\left[ 1-\exp \left( -\frac{z^{2}}{%
\Omega X_{ij}^{2}}\right) \right] ,\quad n<0
\end{equation}%
For $n\geq 0$, $f_{Z}(z)=0$ because positive noise cannot lead to $Y<0$
given $\beta \geq 0$.

\section{Proof of Theorem \ref{theorem:sec_mom}}
\label{appndx-B}

\section*{Complete Solution for the Second Moment of $W_R$}

The second moment is defined as $\mathbb{E}[W_{R}^{2}]=\int_{-\infty }^{\infty
}W_{R}^{2}f_{W}(w_{R})dw_{R}$, where $f_{W_R}(w_R)$ is given in %
\eqref{eq:f_W_R_QPSK_1} and \eqref{eq:f_W_R_QPSK_2}. The overall integral
can be decomposed into three primary terms, $I_1$, $I_2$, and $I_3$.
Integral $I_1$ corresponds to the case of $w_R \ge 0$. 
For $w_{R}\geq 0$, $f_{W_{R}}(w_{R})=\frac{f_{N_{R}}(w_{R})}{2p_{\mathcal{S}}%
}(1+\mu_{j})$, 
\begin{equation}
I_{1}=\frac{1+\mu_{j}}{2p_{\mathcal{S}}}\int_{0}^{\infty
}w_{R}^{2}f_{N_{R}}(w_{R})\,dw_{R}.
\end{equation}%
Since the integral of $w_{R}^{2}f_{N_{R}}(w_{R})$ over the positive
half for a zero-mean Gaussian is $\frac{\sigma _{\mathrm{n}}^{2}}{2}$, then 
\begin{equation}
I_{1}=\frac{1+\mu _{j}}{2p_{\mathcal{S}}}\left( \frac{\sigma _{\mathrm{n}%
}^{2}}{2}\right) =\frac{\sigma _{\mathrm{n}}^{2}(1+\mu _{j})}{4p_{\mathcal{S}%
}}
\end{equation}


For $w_{R}<0$, the integral can be divided into two integrals, $I_2$ and $I_3$, where  
\begin{equation}
I_{2}=\frac{\mu _{j}}{p_{\mathcal{S}}}\int_{-\infty
}^{0}w_{R}^{2}f_{N_{R}}(w_{R})\Phi (\varpi _{1}w_{R})dw_{R}.
\end{equation}
Using the auxiliary function $\Psi (\sigma _{\mathrm{n}},\varpi
)=\int_{-\infty }^{0}w^{2}f_{N_{R}}(w)\Phi (\varpi w)dw$ yields%
\begin{equation}
I_{2}=\frac{\mu _{j}}{p_{\mathcal{S}}}\Psi (\sigma _{\mathrm{n}},\varpi _{1})
\end{equation}%
where $\varpi _{1}=\frac{\chi _{j}}{\lambda _{i}\mu _{j}}$ and the kernel
$$%
\Psi (a,b)=\frac{a^{2}}{2}\left[ \frac{1}{2}-\frac{ba}{\sqrt{2\pi
(1+b^{2}a^{2})}}\right] $$. 

The last integral,
\begin{equation}
I_{3}=\frac{1}{p_{\mathcal{S}}}\int_{-\infty }^{0}w_{R}^{2}f_{N_{R}}(w_{R})%
\mathrm{e}^{-\frac{w_{R}^{2}}{\lambda _{i}^{2}\Omega }}\Phi (\varpi
_{2}w_{R})dw_{R}
\end{equation}
where $\varpi _{2}=-\frac{\chi _{j}}{\lambda _{i}}$. By combining the
exponent from $f_{N_{R}}(w_{R})$ and $\exp \left( -\frac{w_{R}^{2}}{\lambda
_{i}^{2}\Omega }\right) $, and using the change of variables 
\begin{equation*}
\frac{1}{2\varrho _{i}^{2}}=\frac{1}{2\sigma _{\mathrm{n}}^{2}}+\frac{1}{%
\lambda _{i}^{2}\Omega}\text{, } \varrho_{i}=\sigma _{\mathrm{n}}\sqrt{%
\frac{\lambda _{i}^{2}\Omega }{\lambda _{i}^{2}\Omega +2\sigma ^{2}}}=\sigma
_{\mathrm{n}}\mu _{i}
\end{equation*}
Then $I_3$ can be written as, 
\begin{equation}
I_{3}=\frac{\mu _{i}}{p_{\mathcal{S}}}\int_{-\infty }^{0}w_{R}^{2}\left[ 
\tfrac{1}{\sqrt{2\pi }\varrho _{i}}\mathrm{e}^{-\frac{w_{R}^{2}}{2\varrho
_{i}^{2}}}\right] \Phi (\varpi _{2}w_{R})\,dw_{R}
\end{equation}%
which can also be solved using the auxiliary function $\Psi $,%
\begin{equation}
I_{3}=\frac{\mu _{i}}{p_{\mathcal{S}}}\Psi (\sigma _{\mathrm{n}}\mu
_{i},\varpi _{2}).
\end{equation}
Combining the three integrals yields the final closed form%
\begin{equation}
E[W_{R}^{2}]=\frac{1}{p_{\mathcal{S}}}\left[ \frac{\sigma _{\mathrm{n}%
}^{2}(1+\mu _{j})}{4}+\mu _{j}\Psi (\sigma _{\mathrm{n}},\varpi _{1})+\mu
_{i}\Psi (\sigma _{\mathrm{n}}\mu _{i},\varpi_{2})\right] 
\end{equation}
which proves the theorem.

\bibliographystyle{IEEEtran}
\bibliography{Ref}
\end{document}